\documentclass[aps,prb,reprint, amsmath, amssymb,superscriptaddress]{revtex4-2}
\pdfoutput=1
\usepackage[utf8]{inputenc}
\usepackage{bm}
\usepackage[english]{babel}
\usepackage[T1]{fontenc}
\usepackage{mathrsfs}
\usepackage[retainorgcmds]{IEEEtrantools}
\usepackage{color}
\usepackage{amsfonts}
\usepackage{times,txfonts}
\usepackage{nicefrac}
\usepackage{comment}
\usepackage{ulem}
\usepackage[colorlinks=true,linkcolor=blue,urlcolor=blue,citecolor=magenta,pdfusetitle]{hyperref}

\usepackage{epsfig} 
\usepackage{subfigure}
\usepackage{graphicx}
\usepackage{subcaption}
\graphicspath{{Figures/}}

\usepackage{physics}
\usepackage{amssymb}
\usepackage{siunitx}

\usepackage{xcolor}

\usepackage{subfiles}
\begin{document}

\title{Hamiltonian Benchmark of a Solid-State Spin-Photon Interface for Computation}
\author{Tejas Acharya}
\email{tejas.acharya@u.nus.edu}
\affiliation{Department of Physics, National University of Singapore, 117551 Singapore, Singapore}
\affiliation{Centre for Quantum Technologies, National University of Singapore, 117543 Singapore, Singapore}
\affiliation{MajuLab, CNRS-UCA-SU-NUS-NTU International Joint Research Laboratory}
\author{Loïc Lanco}
\affiliation{Université Paris-Saclay, CNRS, Centre de Nanosciences et de Nanotechnologies, 91120 Palaiseau, France}
\affiliation{Université Paris Cité, Centre de Nanosciences et de Nanotechnologies, 91120 Palaiseau, France}
\author{Olivier Krebs}
\affiliation{Université Paris-Saclay, CNRS, Centre de Nanosciences et de Nanotechnologies, 91120 Palaiseau, France}
\author{Hui Khoon Ng}
\affiliation{Department of Physics, National University of Singapore, 117551 Singapore, Singapore}
\affiliation{Centre for Quantum Technologies, National University of Singapore, 117543 Singapore, Singapore}
\affiliation{MajuLab, CNRS-UCA-SU-NUS-NTU International Joint Research Laboratory}
\author{Alexia Auff\`eves}
\affiliation{Centre for Quantum Technologies, National University of Singapore, 117543 Singapore, Singapore}
\affiliation{MajuLab, CNRS-UCA-SU-NUS-NTU International Joint Research Laboratory}
\author{Maria Maffei}
\email{maria.maffei@univ-lorraine.fr}
\affiliation{Université de Lorraine, CNRS, LPCT, F-54000 Nancy, France}
\begin{abstract}
Quantum interfaces are a cornerstone of photonic quantum computing protocols. They consist of quantum emitters coupled to propagating light pulses, enabling the reversible transfer of information between emitter degrees of freedom, such as energy or spin, and photonic degrees of freedom, including polarization or photon number. In practice, propagating light pulses are multi‑mode wavepackets and can undergo substantial spectral distortions during their interactions with the emitters. This effect introduces an intrinsic source of decoherence that has been largely overlooked in early proposals based on idealized monochromatic fields. Here, we investigate the consequences of this decoherence for three key protocols implemented with a solid-state spin-photon interface: the generation of photon‑number superposition states, the realization of a controlled photon‑photon gate, and the generation of photonic cluster states. Our analysis is based on a full Hamiltonian description of the joint emitter–field dynamics. The model further incorporates realistic features of the quantum dot spin dynamics, including decoherence induced by hyperfine interactions. From this framework, we derive analytical expressions for the resulting fidelity losses. Our results demonstrate a pronounced degradation of performance for the photon‑photon gate, a moderate reduction in the fidelity of linear photonic cluster states, and a negligible impact on the generation of photon‑number superpositions. These findings highlight the critical role of multi‑mode effects in assessing the feasibility and scalability of photonic quantum information protocols.
\end{abstract}
\maketitle
\section{Introduction}
Quantum interfaces are composite systems that feature quantum emitters in interaction with propagating electromagnetic fields. They are fundamental building blocks for quantum communication and distributed quantum computing allowing the mapping of stationary qubits, encoded in the emitters, onto flying qubits, encoded in photons~\cite{Kimble_2008}. Confining the field in one dimension maximizes the effectiveness of the light-matter coupling and hence the interface performance~\cite{Ciccarello_OPN2024}. Spin-photon interfaces (SPIs) feature emitters with an internal spin degree of freedom. Due to conservation of angular momentum, transitions between states with different spin projections are coupled with different polarization modes of the electromagnetic field. Energy gaps can be tuned with magnetic fields, and selection rules can be harnessed to tailor specific field's states. Exploiting these mechanisms, SPIs can be used to implement controlled photon-photon gates~\cite{Bonato2010,hu_deterministic_2008} or clusters of entangled photons~\cite{lindner_proposal_2009,economou_spin-photon_2016,Pichler2017}. Experimental implementations of SPIs range from atomic physics \cite{Rempe2007, thomas2022efficient,yang2021sequential}, ion traps \cite{blatt2012}, to solid-state semiconductor devices~\cite{gao_observation_2012,javadi_spinphoton_2018,schwartz_deterministic_2016,Fioretto2022, huet2025,cogan2023,gundin2025spin}. This article focuses on the latter, inspired by recent experimental progress in this field~\cite{Fioretto2022, huet2025,cogan2023,gundin2025spin}.
\\

Early proposals involving SPIs assumed monochromatic photons~\cite{Bonato2010,lindner_proposal_2009, economou_spin-photon_2016}, yet propagating electromagnetic fields are infinite reservoirs of frequency modes and flying photonic qubits correspond to wavepackets of finite bandwidths. In the interaction with quantum emitters shapes and populations of wavepackets change, this introduces a fundamental decoherence mechanism that one must consider in order to exactly characterize the protocols. Moreover in solid-state SPIs the spin suffers from additional decoherence due to the hyperfine interaction with the nuclear spins of the atoms forming the device~\cite{Overhauser_Imamoglu, Merkulov}. This effect has been neglected in most theoretical proposals and included in experimental analyses only as an effective dephasing term in the spin open evolution.
\\
\\
Here we leverage our capacity to solve the joint Hamiltonian dynamics of the emitter-field system using an approach based on the collision model~\cite{Ciccarello_2017} deriving analytical expressions of light-matter entangled wavefunctions~\cite{maffei_closed, maffei2022_Wigner, Maffei_spin, Maffei2024twoqubits}. Our approach does not involve any loss of information on the spin-photon entanglement nor on the state of the multi-mode propagating field. Our Hamiltonian model also captures realistic features of the dynamics of a quantum dot spin including hyperfine interaction with the nuclear environment and yields spin decoherence after classical averaging over many microscopic configurations. We use this framework to derive analytical expressions for the fidelity losses of three protocols: the generation of arbitrary superpositions of photon-number states~\cite{Somaschi2016, tomm_bright_2021}; the controlled photon-photon gate proposed in~\cite{Bonato2010}; and the generation of photonic clusters proposed in~\cite{lindner_proposal_2009} and recently implemented in~\cite{cogan2023,Fioretto2022, huet2025}. Our results show a relevant fidelity loss for the photon-photon gate, a moderate loss for the linear photonic clusters, and a negligible one for the generation of photon-number superpositions.

\begin{figure}[!ht]
    \centering
    \includegraphics[width = 1\columnwidth]{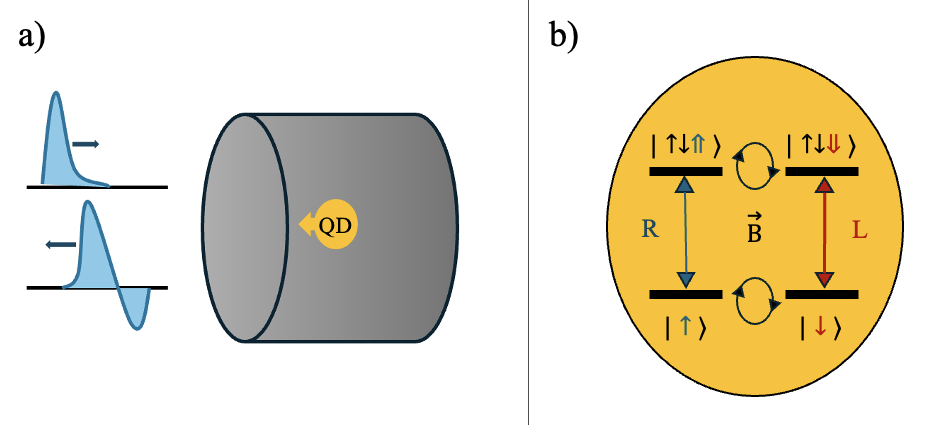}
    \caption{Spin-photon interface implemented with a charged quantum dot (QD) in a directional micro-cavity, schematics. a) The directional micro-cavity confines the field to propagate along half of the $z$ axis, i.e. input and output fields travel on the same side of the emitter. b) Structure of the energy levels of the charged QD: a degenerate 4-level system with optical selection rules, R and L stand for right- and left-circular polarization respectively.}
    \label{figI}
\end{figure}

\section{System and model}\label{secII}

The SPI is implemented with a quantum dot charged with an extra electron spin and embedded in a semi-transparent micro-cavity which provides a highly efficient coupling to the electromagnetic field on one side of the emitter. This structure gives rise to a nearly perfect confinement of the field in one semi-axis, a geometry often called ``half 1D", see Fig.~(\ref{figI}a). All along the paper, we consider that the SPI is kept at a cryogenic temperature and we neglect phonon-induced decoherence and spectral diffusion whose characteristic time scales, in state-of-the-art devices, are considerably longer than the lifetime of the excited states~\cite{Somaschi2016,tomm_bright_2021}.

We consider the configuration where the optical axis, $z$, coincides with the growth direction of the QD. In the absence of an applied magnetic field, the charged QD can be modeled as a degenerate 4-level emitter: two ground states, belonging to the trapped electron, $\left\{\ket{\uparrow},\ket{\downarrow}\right\}$, with spin projections $\pm 1/2$ along $z$; and two excited states, belonging to an optically excited spin-hole pair, or trion, $\left\{\ket{ \uparrow \downarrow \Uparrow},\ket{\uparrow \downarrow \Downarrow}\right\}$, with spin projections along $z$ being $\pm 3/2$. Then conservation of the angular momentum gives rise to optical selection rules involving circular polarizations: the transition $\ket{\downarrow}\rightarrow\ket{\uparrow\downarrow\Downarrow}\left(\text{resp.} \ket{\uparrow}\rightarrow\ket{\uparrow\downarrow\Uparrow}\right)$ can only be promoted by absorption of a left- (resp. right-) circularly polarized photon, see Fig.~(\ref{figI}b). The ground states energies are set equal to zero, such that both trion excited states have energy $\hbar \omega_0$ corresponding to the optical transition frequency. Then the bare Hamiltonian of the QD reads $H_{\text{QD}} = \hbar \omega_0 (\ket{\uparrow \downarrow \Uparrow}\bra{\uparrow \downarrow \Uparrow} + \ket{\uparrow \downarrow \Downarrow}\bra{\uparrow \downarrow \Downarrow})$.  
  
 The propagating field is described as a continuous reservoir of modes of frequency $\omega$ and wavenumber $k$ verifying a linear dispersion relation around the emitter's frequency $\omega_0$, $(k-k_0)=(\omega-\omega_0)/v_g$, with $v_g$ being its group velocity. Adopting the standard treatment for half 1D geometries, we consider only positive values of $k$ and $v_g$ and unfold the $z$ axis around the emitter's position ($z=0$) hence mapping backwards propagation in the negative semi-axis onto forwards propagation in the positive semi-axis. Then, the field's bare Hamiltonian reads $ H_{\text{f}}=\hbar \int d\omega~\omega \left[a^{\dagger}_{R}(\omega)a_{R}(\omega)+a^{\dagger}_{L}(\omega)a_{L}(\omega)\right]$, where $L$ and $R$ stand for left- and right-circularly polarized photons respectively, and $a_{L/R}(\omega)$ are annihilation operators verifying $[a_{j}(\omega),a^{\dagger}_{i}(\omega')]=\delta_{i,j}\delta(\omega-\omega')$ with $i,j\in\lbrace L, R\rbrace$.
 
 The light-matter interaction is treated with the \textit{input-output} formalism describing localized quantum emitters weakly coupled with electromagnetic fields propagating in 1D~\cite{Gardiner1985Input, combes_slh_2017,kiilerich_quantum_2020,Rambeau2026}. This treatment assumes rotating wave approximation and uniform coupling between the emitter and the frequency modes, such that the interaction Hamiltonian reads $V=i\hbar \int d\omega~ \sqrt{\gamma/(2\pi)}\left( \vert\downarrow\rangle\langle\uparrow \downarrow \Downarrow\vert \otimes a_{L}^{\dagger}(\omega) +\vert\uparrow\rangle\langle\uparrow \downarrow \Uparrow \vert \otimes a_{R}^{\dagger}(\omega)\right)-h.c.$. In this regime, we can define quantum noise operators destroying photons in the position $z$ at time $t$: $b_{R(L)}(t,z)= (2\pi)^{-1/2} \int d\omega e^{-i (\omega-\omega_0)(t - z/vg )}a_{R(L)}(\omega)=b_{R(L)}(t-z/vg,0)$, with $[b_{i}(t,0),b^{\dagger}_{j}(t',0)]=\delta_{i,j}\delta(t-t')$~\cite{Gardiner-Zoller-book}. The QD is positioned in $z=0$, so, from now on we will set $b_{j}(t,0)\equiv b(t)$ to shorten the notation. Then, in the interaction picture with respect to $H_{\text{QD}}+H_{\text{f}}$, the light-matter coupling becomes
\begin{align}\label{eq:V_continuousTime}
V_{\text{I}}(t)=i \hbar  \sqrt{\gamma}~\left(\vert\downarrow\rangle \langle \uparrow \downarrow \Downarrow\vert \otimes b_{L}^{\dagger}(t) +\vert \uparrow\rangle \langle\uparrow \downarrow \Uparrow\vert \otimes b_{R}^{\dagger}(t)\right)-h.c.
\end{align}

\subsection{Magnetic Hamiltonian}

In a realistic QD at low temperatures and weak magnetic field (few tens of mT) the spin undergoes two kinds of dynamics: a coherent precession induced by an external magnetic field, and a decoherence process due to the interaction with the surrounding nuclei. Here we consider only the so-called Voigt configuration where the external magnetic field is perpendicular to $z$. A complete description should consider that  the trion Land\'e factor is a tensor, such that the magnetic Hamiltonian of the trion spin would depend on the magnetic field orientation. However it is possible to find a direction of the external magnetic field, in the plane perpendicular to $z$, giving rise to a magnetic Hamiltonian whose quantization axis is parallel to the magnetic field itself as if the trion Land\'e factor would be just a positive scalar, $g_{\text{tr}}$. Such specific configuration, recently demonstrated feasible in experiments~\cite{Krebs2025, Serov2025}, is the one considered in the ideal protocols of Sec.~\ref{results}. In the rest of the paper we will assume this optimal configuration. We take the direction of the external magnetic field as our $x$ axis, $\textbf{B}^{\text{ext}}=B^{\text{ext}}\hat{x}$, and then we write the magnetic Hamiltonian as $H_{\text{s}}^{\text{ext}}=\mu_B B^{\text{ext}}\left[g_{\text{tr}}s_x^\text{tr} +g_{\text{el}} s_x^\text{el}\right]/2$, where $g_{\text{el}}$ is the electron Land\'e factor, $\mu_B$ is the Bohr magneton, and $s_x^\text{tr(el)}\equiv s^{\text{tr(el)}}+ (s^{\text{tr(el)}})^{\dagger}$ are trion and electron $x$ Pauli matrices, with $s^{\text{tr}}=\ket{\uparrow \downarrow \Downarrow}\bra{\uparrow \downarrow \Uparrow}$ and $s^{\text{el}}=\ket{\downarrow}\bra{\uparrow}$. Let us notice that, due to the optical selection rules, $H_{\text{s}}^{\text{ext}}$ and and $V$ do not commute. In this situation, a no-go theorem of quantum foundations, known as WAY theorem~\cite{Loveridge2011Measurement}, applies and states that it is impossible to obtain a perfect one-to-one spin-photon mapping aside of the limit $g_{\text{tr(el)}}\mu_B B^{\text{ext}}/(\hbar\gamma)\rightarrow 0$ where spin precession and light-matter interaction become effectively separable. Hence ideal protocols would require $g_{\text{tr}}=g_{\text{el}}=g$ and $g\mu_B B^{\text{ext}}/(\hbar\gamma)= 0$. As we will show in Sec.~\ref{results}, the unavoidable deviations from these conditions, included in our model, lower protocols fidelity.
\\
At low temperatures, the main incoherent contribution to the spin dynamics comes from the hyperfine interaction of the electron spin with the nuclear spins of the $\sim 10^{4}$-$10^{6}$ atoms that compose the nanostructure, this interaction can be modeled as an effective magnetic field, called Overhauser field, $\textbf{B}^{\text{O}}$, around which the electron spin effectively precesses~\cite{Overhauser_Imamoglu}. Let us notice that also the trion spin is effectively coupled with an Overhauser field, but since the latter is one order of magnitude smaller we neglect it. Notice that this approximation limits the accuracy of the fidelity results in Sec.~\ref{results} to $\sim 10^{-3}$. The intensity of the (electron) Overhauser field is of the order of tens of mT and its direction fluctuates as result of the changing nuclear environment. The most important semi-classical model to describe the electron spin decoherence in semiconductor QDs is given by the Merkulov-Efros-Rosen model~\cite{Merkulov}, in which three distinct stages are described, each with its own characteristic timescale. In the first stage, or \textit{frozen-nuclear-spin stage}, of the duration of a few hundred nanoseconds, the Overhauser field is considered static so that the electron spin coherently precesses around it. The statistical averaging over all possible intensities and orientations of the Overhauser field leads to spin decoherence on a few nanoseconds timescale~\cite{Merkulov, Bechtold2015}, until reduction of the average electronic spin polarization by a factor of $1/3$ with respect to its initial value. The model then predicts a second stage where the Overhauser field is no longer static leading to a further reduction by a factor $1/3$, until complete relaxation occurs after a few hundred microseconds. In this paper, we study dynamics taking place on time-scales of a few hundred of nanoseconds meaning within the first stage of the relaxation process. We then consider the Overhauser field frozen in a certain configuration and we add its coupling with the electron spin to the magnetic Hamiltonian of the external field:
\begin{align}\label{eq_Hs}
H_{\text{s}} & = H_{\text{s}}^{\text{ext}}+ \frac{\mu_B g_{\text{el}}}{2}\textbf{B}^{\text{O}}\cdot \textbf{s}^\text{el} \equiv \frac{\hbar}{2}\left( \Omega_g\textbf{n}\cdot \textbf{s}^\text{el} + \Omega_e s_{x}^\text{tr} \right)
\end{align}
where we set $\hbar\Omega_e=g_{\text{tr}}\mu_B B^{\text{ext}}$ and $\hbar \Omega_{g}\textbf{n}=g_{\text{el}}\mu_B \left(\textbf{B}^{\text{O}}+\textbf{B}^{\text{ext}}\right)$, with $\textbf{n}=\left(\sin{(\theta)}\cos{(\phi)}; \sin{(\theta)}\sin{(\phi)};\cos{(\theta)}\right)$ being a unit vector and $\textbf{s}^\text{el}=(s_x^\text{el}; s_y^\text{el}; s_z^\text{el})$ being the vector of spin Pauli matrices for the electron spin. The expectation values of the observables are then computed by taking first quantum mechanical averages on pure states, corresponding to fixed Overhauser field configurations, and then classical averages over all possible configurations. The latter are performed by using an isotropic Gaussian distribution of the Overhauser field configurations centered around zero and having standard deviation $\Delta_{\text{B}^{\text{O}}}=\hbar/(g_{\text{el}}\mu_{B}) w$, where $w$ is the inverse of the electron spin coherence time, $w^{-1}=T_2^*\sim 1$ ns~\cite{Merkulov, Bechtold2015}. So, for instance, the expectation value of the electron spin polarization, $s_z^{\text{el}}$, in the absence of external magnetic field, i.e. $\Omega_{e}=0$ and $\Omega_{g}=\Omega_{\text{O}}=g_{\text{el}}~\mu_B |\textbf{B}^{\text{O}}|/\hbar$, starting from the electron spin state $\ket{\uparrow}$, is obtained from the statistical average of $\langle s_{z}^\text{el}\rangle=\cos^2\left(\frac{\Omega_{\text{O}} t}{2}\right) +\sin^2\left(\frac{\Omega_{\text{O}} t}{2}\right)\cos{(2\theta)}$: 
\begin{align}\label{eq_Merkulov}
&\overline{ s_z^{\text{el}}}=\frac{\int_{0}^{\infty} \Omega_{\text{O}}^2~d\Omega_{\text{O}}\int_{0}^{\pi} \sin(\theta)~d\theta  \int_{0}^{2\pi} d\phi~e^{-\Omega_{\text{O}}^2/2w^2}\langle s_z^{\text{el}}\rangle}{(2\pi w^2)^{3/2}}\\ \nonumber
&=\frac{1}{3}\left( 1 + 2 e^{-w^2 t^2/2}(1 - t^2 w^2)\right),
\end{align}
which, in perfect agreement with the semi-classical prediction~\cite{Merkulov}, reduces to $1/3$ for $t\gg T_{2}^{*}=w^{-1}$. In the rest of the paper, we use the notation $\overline{X}$ to denote statistical averages of quantum mechanical expectation values that in turn will be denoted as $\langle X\rangle$.

\subsection{Collision model of the SPI dynamics}

Let us introduce a new notation that will help us simplifying the calculations by decoupling the energy and the spin degrees of freedom. The basis of the SPI energy levels gets transformed as: $\left\{\ket{ \uparrow},\ket{\downarrow}\right\}\rightarrow \left\{\ket{\uparrow}\ket{g},\ket{\downarrow}\ket{g}\right\}$ and $\left\{\ket{ \uparrow \downarrow \Uparrow},\ket{\uparrow \downarrow \Downarrow}\right\}\rightarrow \left\{\ket{\uparrow}\ket{e},\ket{\downarrow}\ket{e}\right\}$. In the new basis, we define ladder operators acting on spin and energy degrees of freedom: $s \equiv\vert\downarrow\rangle\langle\uparrow\vert,~\sigma\equiv\vert g\rangle\langle e\vert$, $s_{z}\equiv\vert\uparrow\rangle\langle\uparrow\vert - \vert\downarrow\rangle\langle\downarrow\vert$, and $\sigma_{z}\equiv\vert e\rangle\langle e\vert - \vert g\rangle\langle g\vert$. We will dub this notation "decoupled notation", and we will use it in all the equations from now on, unless explicitly stated otherwise. In the decoupled notation, omitting all identity operators, the terms of the Hamiltonian read: $H_\text{QD} = \hbar\omega_0 \sigma^\dagger \sigma$ ; $V_{\text{I}}(t) = i\hbar \sqrt{\gamma} (s s^\dagger \otimes \sigma \otimes b^\dagger_L(t) + s^\dagger s \otimes \sigma \otimes b_R^\dagger(t)) -h.c$; and $H_{\text{s}}=\hbar/2\left( \Omega_{g}  \textbf{n}\cdot\textbf{s} \otimes \sigma\sigma^{\dagger}+ \Omega_{e} s_x \otimes \sigma^{\dagger}\sigma\right)$.
\\
\\
We are now ready to solve the SPI dynamics using an approach based on collision model~\cite{Ciccarello_2017} that gives analytical expressions of the joint emitter-field wavefunctions as shown in Refs.~\cite{maffei_closed,Maffei_spin, maffei2022_Wigner, Maffei2024twoqubits}. We take a coarse-graining time $\Delta t\ll\gamma^{-1}$, and define quantum noise increment operators $b_{j,n}\equiv (\Delta t)^{-1/2}\int_{n \Delta t}^{(n+1)\Delta t} dt b_{j}(t)$ with $j=R,L$ and $[b_{i,n},b^{\dagger}_{j,m}]=\delta_{i,j}\delta_{n,m}$. The operator $b_{R(L),n}$ destroys excitations in the $n$-th time-bin mode, or collision unit, of the right(left)-circularly polarized field that now is regarded as a sequence of units passing by the emitter's position on a conveyor belt. Using the commutation relations of the operators $b_{j,n}$ at different $n$, the unitary evolution operator evolving the joint light-matter system during the time interval $[n\Delta t,(n+1)\Delta t]$ can be written explicitly in the exponential form:
\begin{equation}\label{eq_Un}
    U_n =\text{exp}\left\lbrace- \frac{i}{\hbar}\Delta t (H_{\text{s}} + V_n) \right\rbrace
\end{equation}
with 
\begin{align}\label{eq:V_discreteTime}
&V_n\equiv \frac{1}{\Delta t}\int_{n \Delta t}^{(n+1)\Delta t}dt' V_{\text{I}}(t')\\ \nonumber
&=\hbar i \sqrt{\frac{\gamma}{\Delta t}}~ \left( ss^\dagger \otimes \sigma \otimes b_{L,n}^{\dagger} + s^\dagger s \otimes \sigma\otimes b_{R,n}^{\dagger}\right)-h.c. 
\end{align}
It follows that the evolution until any time $t$ can be decomposed in a product of $M=t/\Delta t$ independent collisions, i.e. $\ket{\Psi(t)}=\text{lim}_{\Delta t \rightarrow 0} \prod_{n=0}^{M-1}U_n\ket{\Psi(0)}$. 
Writing the vacuum of the electromagnetic field in the basis of the collision units, $\ket{0_R,0_L}=\bigotimes_{n} \ket{0_{R,n},0_{L,n}}$, with $b_{j,n}\ket{0_{j,n}}=0$ and $b^{\dagger}_{j,n}\ket{0_{j,n}}=\ket{1_{j,n}}$, and expanding $U_n$ up to the first order in $\gamma\Delta t$, one finds compact expressions for the emitter's Kraus operators corresponding to its no-jump evolution: 
\begin{align}\label{eq_K_no_j}
\mathcal{K}(t)\equiv \mathrm{lim}_{\Delta t\rightarrow 0}\left(\Pi_{n=0}^{t/\Delta t}\bra{0_{R,n},0_{L,n}}U_n \ket{0_{R,n},0_{L,n}}\right)\\ \nonumber = e^{-i\frac{\Omega_g t}{2}\textbf{n}\cdot\textbf{s}}\otimes \sigma \sigma^\dagger+ e^{-\gamma t/2} ~e^{-i\frac{\Omega_e t}{2} s_x}\otimes \sigma^\dagger \sigma,
\end{align}
to its jump operators:
\begin{align}\label{eq_K_min}
\mathcal{J}^{(-)}_{R}\equiv \mathrm{lim}_{\Delta t\rightarrow 0}\frac{\bra{1_{R,n}, 0_{L,n}} U_{n} \ket{0_{R,n},0_{L,n}}}{\sqrt{\Delta t}}=\sqrt{\gamma}  s^\dagger s \otimes \sigma,\\ \nonumber \mathcal{J}^{(-)}_{L}\equiv \mathrm{lim}_{\Delta t\rightarrow 0}\frac{\bra{0_{R,n}, 1_{L,n}} U_{n} \ket{0_{R,n},0_{L,n}}}{\sqrt{\Delta t}}=\sqrt{\gamma} s s^\dagger \otimes \sigma,
\end{align}
and to its absorption operators:
\begin{align}\label{eq_K_plus}
\mathcal{J}^{(+)}_{R}\equiv \mathrm{lim}_{\Delta t\rightarrow 0}\frac{\bra{0_{R,n}, 0_{L,n}} U_{n} \ket{1_{R,n},0_{L,n}}}{\sqrt{\Delta t}}=-\sqrt{\gamma}  s^\dagger s \otimes \sigma^{\dagger},\\ 
\mathcal{J}^{(+)}_{L}\equiv \mathrm{lim}_{\Delta t\rightarrow 0}\frac{\bra{0_{R,n}, 0_{L,n}} U_{n} \ket{0_{R,n},1_{L,n}}}{\sqrt{\Delta t}}=-\sqrt{\gamma}  s s^\dagger \otimes \sigma^{\dagger}. \nonumber
\end{align}
In the following we use the above operators to derive analytical expressions of the SPI wavefunctions with different initial conditions.

\section{Results}\label{results}

Here we apply the Hamiltonian model described in the previous section to three protocols: generation of photon-number superpositions (\ref{PNS}), photon-photon gate (\ref{CZ}) and Lindner-Rudolph protocol for cluster states generation (\ref{LR}). For all protocols we first describe the ideal dynamics (yielding unitary fidelity) as presented in the original theoretical proposals using monochromatic field's description and neglecting every source of decoherence. Then, using our model, we solve the dynamics with multi-mode description of the propagating field, finite ratio $g_{\text{tr(el)}}\mu_B B^{\text{ext}}/(\hbar \gamma)$, unequal Land\'e factors $g_{\text{tr}}\neq g_{\text{el}}$, spin decoherence due to hyperfine interaction. We neglect imperfections in the alignment of the external magnetic field with the trion's quantization axis, as well as pure dephasing and spectral fluctuations. This choice allows us to derive analytical results that single out the impact of the decoherence mechanisms that we consider. Due to our assumptions, the obtained fidelity values must be interpreted as upper bounds of those experimentally achievable with state-of-the-art devices.

\subsection{Generation of photon-number superpositions}\label{PNS}
In this section, we benchmark the mapping of a qubit encoded in the QD's energy onto a flying qubit encoded in the photon number: 
\begin{align}\label{eq:map1}
\left(\alpha\ket{e}+\beta\ket{g}\right)\ket{0}\rightarrow \ket{g}\left(\alpha\ket{1}+\beta\ket{0}\right).
\end{align} 
The energy qubit can be prepared using a classical pulse, and with the ideal map above, one can prepare any such superposition of 0- and 1-photon number states. Such a map does not require any external magnetic field, and in the ideal case, where the Overhauser field is also negligible, the action of the sole light-matter coupling for a time $t_{\infty}\gg \gamma^{-1}$ would give exactly Eq.~\eqref{eq:map1} with $ \ket{1}$ being either $\ket{1_{R}}=\sqrt{\gamma} \int_{0}^{\infty} dt'  e^{-\gamma t'/2 }b^{\dagger}_{R}(t')\ket{0}$, or $\ket{1_{L}}=\sqrt{\gamma} \int_{0}^{\infty} dt'  e^{-\gamma t'/2 }b^{\dagger}_{L}(t')\ket{0}$, according to the SPI's initial state being $\ket{\uparrow}\otimes \left(\alpha\ket{e}+\beta\ket{g}\right)$ or $\ket{\downarrow}\otimes \left(\alpha\ket{e}+\beta\ket{g}\right)$. In the following, without loss of generality, we'll take the initial state $\ket{\Psi^{(\uparrow)}_{0}}=\ket{\uparrow}\otimes \left(\alpha\ket{e}+\beta\ket{g}\right)$, and hence $\ket{1}=\ket{1_{R}}$. We set $\Omega_e=0$ and $\Omega_g=\Omega_{\text{O}}$ in the magnetic Hamiltonian $H_{\text{s}}$ and compute the evolution of the SPI under the repeated action of the collision unitary [Eq.~\eqref{eq_Un}] finding its state at time $t_{\infty}\gg \gamma^{-1}$ ($e^{-\gamma t_{\infty}/2}\approx 0$):
\begin{align}\label{eq_real_state1}
\ket{\Psi^{(\uparrow)}_{\infty}}=&\left[\alpha\sum_{\mu=\uparrow,\downarrow}\int_{0}^{\infty} dt' f_{1_{R}}^{(\uparrow,\mu)}(t_{\infty},t')b^{\dagger}_{R}(t')\ket{\mu} \right.\\\nonumber
&\left. + \beta~e^{-i\frac{\Omega_{\text{O}} t_{\infty}}{2}\textbf{n}\cdot\textbf{s}}\ket{\uparrow}\right]\ket{0,g},
\end{align}
with
\begin{align}
&f_{1_{R}}^{(\uparrow,\uparrow)}(t,t^{\prime})=\bra{g,\uparrow}\mathcal{K}(t-t')\mathcal{J}^{(-)}_{R}\mathcal{K}(t')\ket{\uparrow,e} \\ 
&=\sqrt{\gamma}e^{-\gamma t^{\prime}/2} \left(\cos\left(\frac{\Omega_{\text{O}}(t-t^{\prime})}{2}\right)-i\cos{(\theta)}\sin\left(\frac{\Omega_{\text{O}}(t-t^{\prime})}{2}\right)\right)\nonumber
\end{align}
and
\begin{align}
&f_{1_{R}}^{(\uparrow,\downarrow)}(t,t^{\prime})=\bra{g,\downarrow}\mathcal{K}(t-t')\mathcal{J}^{(-)}_{R}\mathcal{K}(t')\ket{\uparrow,e} \\ \nonumber
&= -i\sqrt{\gamma}e^{i\phi}\sin{(\theta)}e^{-\gamma t^{\prime}/2} \sin\left(\frac{\Omega_{\text{O}}(t-t^{\prime})}{2}\right),
\end{align}
 where the operators $\mathcal{K}$ and $\mathcal{J}^{(-)}_{R}$ are given in Eq.~\eqref{eq_K_no_j} and Eq.~\eqref{eq_K_min} and the explicit expressions of all coefficients $f_{j}^{(\zeta,\mu)}(t,t')$ are given in Appendix~\ref{App:A}. Notice that, due to the Overhauser field, $\ket{\Psi_{\infty}^{(\uparrow)}}$ is an entangled spin-photon state. Its generation can be interpreted as follows: a spontaneous emission event (jump) turns on the precession of the electron spin due to the Overhauser field, since the jump can occur at any time between $0$ and $t_{\infty}$, the wavefunction is a coherent superposition of jumps occurring at different times followed by electron spin precession starting at that time.
 
\begin{figure}[h]
    \centering
    \includegraphics[width=1\linewidth]{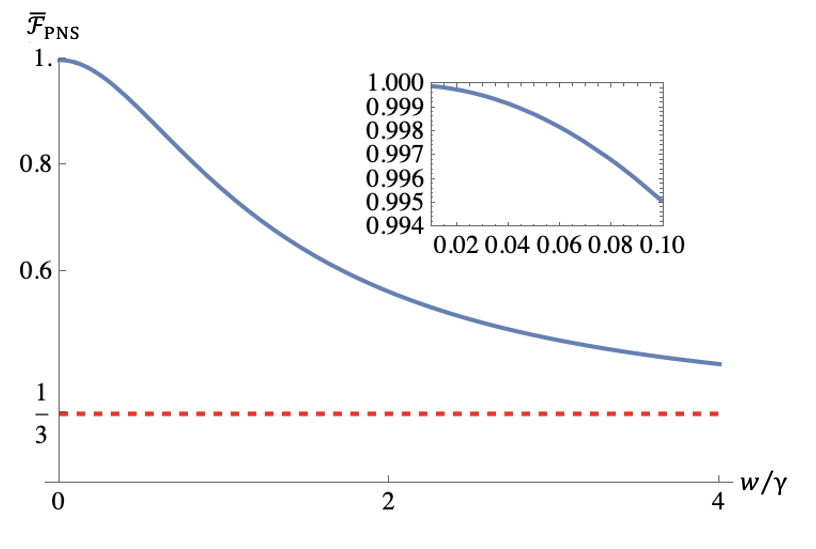}
    \caption{Fidelity upper bound of the superposition of 0- and 1-photon states varying the spin coherence time $T_{2}^{*}=w^{-1}$ with respect to the trion's lifetime $\gamma^{-1}$. The plotted function is given in Eq.~\eqref{eq_final_photon_source}. State of the art QD devices having $w/\gamma \sim 0.1$ correspond to a nearly unitary fidelity upper bound (see the inset).}
    \label{figII}
\end{figure}

 Then, to compare the performance against the ideal map [Eq.~\eqref{eq:map1}], we first take the trace over the spin degree of freedom, obtaining the partially mixed state $\text{Tr}_{\text{spin}} \left[\ket{\Psi_\infty^{(\uparrow)}}\bra{\Psi_\infty^{(\uparrow)}} \right] \equiv \rho_{\infty}$. We then compute its fidelity with the ideal state appearing in the right member of Eq.~\eqref{eq:map1}:
 \begin{align}\label{pns_fid}
     \mathcal{F}_{\text{PNS}}(\alpha,\beta) & = \bra{\Psi_\text{ideal}} \rho_\infty\ket{\Psi_\text{ideal}} \nonumber \\ 
     &= \frac{\abs{\alpha}^4 + 2\abs{\alpha}^2\abs{\beta}^2}{1 + \left(\frac{\Omega_{\text{O}}}{2\gamma}\right)^2} + \abs{\beta}^4,
 \end{align}
 where PNS stands for photon number superposition. The above expression is explicitly derived in Appendix~\ref{App:B}. Note that, if we started from the state $\ket{\downarrow}\otimes \left(\alpha\ket{e}+\beta\ket{g}\right)$, hence obtaining a left-circularly polarized photon instead of a right-circularly polarized one, the fidelity would have been the same, as shown in Appendix~\ref{App:B}. To get rid of the fidelity's dependence on the choice of the parameters $\alpha$ and $\beta$, we can integrate $\mathcal{F}_{\text{PNS}}(\alpha,\beta)$ over the uniform measure of the energy qubit's Bloch sphere~\cite{Nielsen_2002}, obtaining:
\begin{align}\label{eq_prelimin_photon_sorce}
   \mathcal{F}_{\text{PNS}}= \frac{1 + \left(\frac{\Omega_{\text{O}}}{2\sqrt{3}\gamma}\right)^2}{1 + \left(\frac{\Omega_{\text{O}}}{2\gamma}\right)^2}.
\end{align}
Taking the statistical average of the above fidelity with the isotropic Gaussian distribution of standard deviation $w$, we obtain: 
\begin{align}\label{eq_final_photon_source}
    \overline{\mathcal{F}}_{\text{PNS}}&=\sqrt{\frac{2}{\pi w^{6}}}\int_{0}^{\infty} d\Omega_{\text{O}}~ e^{-\Omega_{\text{O}}^2/2 w^2} \mathcal{F}_{\text{PNS}} \\ \nonumber
    &=\frac{8(w/\gamma) + (w/\gamma)^3 - 8 \sqrt{2 \pi}~e^{\frac{2}{(w/\gamma)^2}}~\text{Erfc}\left[\frac{\sqrt{2}}{(w/\gamma)}\right]}{3(w/\gamma)^3}.
\end{align}
Let us notice that, in order to compute the limit of $\overline{\mathcal{F}}_{\text{PNS}}$ for $x = w/\gamma \rightarrow 0$, it is convenient to replace the last term with its Taylor expansion around $x=0$, obtaining $8\exp{2/x^2} \sqrt{2\pi}\text{Erfc}[\sqrt{2}/x] = 8x - 2x^3 + 3/2 x^5 + O(x^6)$ that gives $\overline{\mathcal{F}}_{\text{PNS}}=1-x^2/2 + O(x^3)$. As shown in Fig.~(\ref{figII}), the fidelity upper bound goes to $1$ in the limit of infinite electron spin coherence time, $w/\gamma \rightarrow 0$, and to $1/3$ in the opposite limit, where the electron spin relaxation is so fast that the average polarization reaches the value of $1/3$ before the spontaneous emission even begins. As mentioned in the previous section, the spin coherence time, $T_{2}^{*}=w^{-1}$, is typically of a few nanoseconds. On the other hand, QDs embedded in high-quality micro-cavities at cryogenic temperature (as for instance in Refs.~\cite{Somaschi2016, tomm_bright_2021, wein2022photon, Fioretto2022}), have excited states lifetimes, $\gamma^{-1}$, that can reach a few hundred picoseconds. Hence, a realistic value of the adimensional ratio $w/\gamma$ is about $0.1$. As shown in Fig.~(\ref{figII}), for values of $w/\gamma$ between 0.01 and 0.1, the fidelity upper bound is above $99\%$. We then conclude that electron spin decoherence does not affect relevantly the generation of 0- and 1-photon states superpositions.   

\subsection{Photon-photon gate}\label{CZ}

The simplest design of controlled-Z (CZ) photon-photon gate, as proposed in Ref.~\cite{Bonato2010}, exploits the phase shift acquired by a resonant photon in the scattering with a two-level atom. In the ideal case, of monochromatic photons, the phase acquired in the scattering is exactly $\pi$. This mechanism, in the SPI, gives rise to a spin-phase gate, $\mathcal{H}$, controlled by the presence of one circularly polarized photon:
\begin{align}\label{eq_phase_gate}
&\ket{\uparrow}\ket{g}\ket{1_R}\xrightarrow{\mathcal{H}}-\ket{\uparrow}\ket{g}\ket{1_R},\\ \nonumber
&\ket{\downarrow}\ket{g}\ket{1_R}\xrightarrow{\mathcal{H}}\ket{\downarrow}\ket{g}\ket{1_R},
\end{align}
and analogously for a left polarized photon exchanging the roles of $\ket{\uparrow}$and $ \ket{\downarrow}$. Then, as in the previous section, without loss of generality, from now on we will take the logical basis of the photons to be $\ket{1_R}=\ket{1}$ and $\ket{0}=\ket{0}$. 

\begin{figure}
    \centering
    \includegraphics[width=1\linewidth]{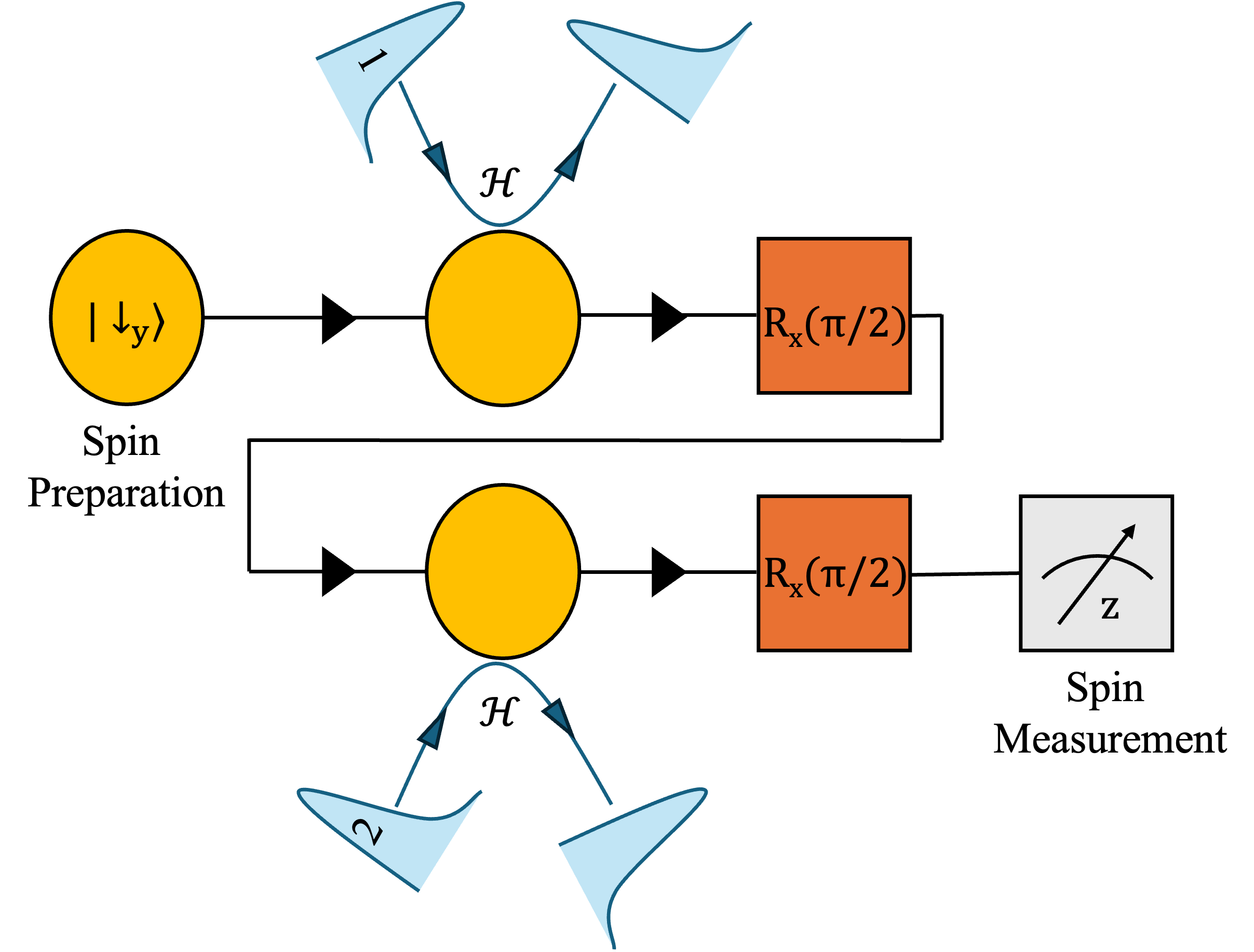}
    \caption{Schematic representation of the ideal CZ photonic gate. The electron spin is prepared in $\ket{\downarrow_y}= \left(\ket{\uparrow} -i \ket{\downarrow}\right)/\sqrt{2}$. Both the control and the target photon are prepared in arbitrary superpositions of $0-$ and $1-$photon states. The scattering of each photon corresponds to the spin-selective phase map, $\mathcal{H}$, of Eq.~\eqref{eq_phase_gate}. After each scattering, the electron spin is rotated around the $x$ axis of $\pi/2$. The spin-photon-photon state is given at this stage by Eq.~\eqref{ideal_cz}. The electron spin is then measured, projecting the state onto a CZ gate up to a single photon unitary (see text).}
    \label{CZSchematic}
\end{figure} 

 Let us briefly recall the ideal protocol which is schematically represented in figure Fig.~(\ref{CZSchematic}). The electron spin is prepared in $\ket{\downarrow_y} =\left( \ket{\uparrow} -i \ket{\downarrow}\right)/\sqrt{2}$; then the first R-photon, namely the control qubit, is sent in the SPI which scatters it according to Eq.~\eqref{eq_phase_gate}; then a $\pi/2$ rotation around the $x$ axis, $R_{x}(\pi/2)$, is applied on the electron spin:
\begin{align}
\ket{\downarrow_y}(\alpha_1 \ket{0}_1 + \beta_1\ket{1}_1)
&\xrightarrow{R_{x}(\pi/2) \mathcal{H}_1} -(i\alpha_1\ket{0}_1 \ket{\downarrow} + \beta_1\ket{1}_1\ket{\uparrow}),
\end{align}
then the second R-photon, namely the target, is sent in the SPI followed by a second $R_{x}(\pi/2)$ rotation:
\begin{align}\label{ideal_cz}
&\ket{\downarrow_y}(\alpha_1\ket{0}_1 + \beta_1\ket{1}_1)(\alpha_2\ket{0}_2 + \beta_2\ket{1}_2)\xrightarrow{R_{x}(\pi/2)\mathcal{H}_2 R_{x}(\pi/2) \mathcal{H}_1} \nonumber \\
&-\frac{1}{\sqrt{2}} \ket{\uparrow} [\alpha_1 \alpha_2 \ket{0}_1\ket{0}_2 + \alpha_2\beta_1 \ket{1}_1\ket{0}_2 \nonumber\\ 
&+\alpha_1\beta_2\ket{0}_1\ket{1}_2 - \beta_1\beta_2 \ket{1}_1\ket{1}_2] \nonumber \\
&-\frac{i}{\sqrt{2}}\ket{\downarrow}[\alpha_1 \alpha_2 \ket{0}_1\ket{0}_2 - \alpha_2\beta_1 \ket{1}_1\ket{0}_2  \nonumber \\
&+\alpha_1\beta_2\ket{0}_1\ket{1}_2 + \beta_1\beta_2 \ket{1}_1\ket{1}_2]. 
\end{align}
Let us notice that, in this ideal protocol, each photon scattering corresponds to the application of a spin-phase gate $\cal{H}$ [Eq.~\eqref{eq_phase_gate}], subscripts 1 and 2 referring to first and second photon. At the end of this sequence, the electron spin is measured in the $z$ direction. If it is found in $\ket{\uparrow}$, the photons have undergone the CZ gate, if it is found in $\ket{\downarrow}$, a $Z$ gate must be applied to the control photon. In the following, for simplicity, we will compute the fidelity of the gate in the case where the final spin measurement projects it on $\ket{\uparrow}$.
\\
\\
In a realistic implementation, the spin cannot be manipulated unitarily between the single-photon scatterings, as no external magnetic field could be turned on and off instantaneously. Thus, the realistic protocol entails: turning on a transverse magnetic field, preparing the spin in the state $\ket{\downarrow_y}$, sending the control photon to the emitter (while the magnetic field is still on), letting the system evolve for a time $T_g=\pi/(2\Omega_g)$, then sending the target photon, let the SPI evolve for another time interval $T_{g}$. Notice that in the following we assume that the scattering of the first photon is completed before the second photon arrives in the SPI. 
We then look for the final SPI state in the long-time limit (when the scattering is completed), starting from the initial states $\ket{\chi^{(\zeta)}_0}=\ket{\zeta,g,1_{R}}$, with $\zeta=\uparrow,\downarrow$ and $\ket{1_{R}}=\int_{0}^{\infty} dt' \xi_{R}(t') b_{R}^{\dagger}(t')\ket{0}$ and $\int_{0}^{\infty} dt' |\xi_{R}(t')|^2=1 $. Using the Kraus operators [Eqs.~\eqref{eq_K_no_j}-\eqref{eq_K_plus}] we obtain:
\begin{align}\label{eq_scattering}
\ket{\chi^{(\zeta)}_\infty}=
\sum_{\mu = \uparrow, \downarrow}\sum_{j=R,L}\int_{0}^{t_\infty}dt' \lambda_{j}^{(\zeta,\mu)}(t,t') b_{j}^{\dagger}(t')\ket{0} \ket{\mu}\ket{g}
\end{align}
with
\begin{align}\label{eq_WF1_scatteringR}
&\lambda_{R}^{(\zeta,\mu)}(t,t') = \xi(t') \bra{g,\mu} \mathcal{K}(t)\ket{\zeta, g} \nonumber \\
+ &\int_0^{t'} dt'' \xi(t'') \bra{g,\mu} \mathcal{K}(t-t')\mathcal{J}^{(-)}_R \mathcal{K}(t' - t'')\mathcal{J}_R^{(+)} \mathcal{K}(t'')\ket{\zeta, g}
\end{align}
and
\begin{align}\label{eq_WF1_scatteringL}
&\lambda_{L}^{(\zeta,\mu)}(t,t') \nonumber \\&= \int_{0}^{t'} dt'' \xi(t'')\bra{g,\mu}\mathcal{K}(t-t')\mathcal{J}^{(-)}_{L}\mathcal{K}(t'-t'')\mathcal{J}_{R}^{(+)}\mathcal{K}(t'')\ket{\zeta,g},
\end{align}
the explicit expressions of the coefficients $\lambda_R^{(\zeta, \mu)}(t, t')$ are given in Appendix~\ref{App:A}. The assumption that the scattering of the first photon is completed before arrival of the second, implies that photons 1 and 2 have support on different time-bins modes, resp. $[0,T_{g}]$ and $[T_{g},2T_{g}]$, and we just need to combine two copies of 
Eq.~\eqref{eq_scattering} to get the SPI state at the end of the realistic protocol, $t=2T_{g}$:
\begin{align}
&\ket{\Phi(2T_g)}= -\frac{\alpha_1\alpha_2}{\sqrt{2}}\ket{0}_1\ket{0}_2 (C_{-} \ket{\uparrow} +i C_{+}\ket{\downarrow}) \\ \nonumber
& + \frac{\alpha_2\beta_1}{\sqrt{2}} \ket{0}_2\sum_{j, \mu}(\ket{\omega_j^{(\uparrow, \mu)}}_1 - i \ket{\omega_j^{(\downarrow, \mu)}}_1)e^{\frac{-i\pi \mathbf{n\cdot s}}{4}}\ket{\mu} \\ \nonumber
& + \frac{\alpha_1\beta_2}{2} \ket{0}_1\sum_{k, \nu} [(1 -C_{-})\ket{\omega_k^{(\uparrow, \nu)}}_2 \ket{\nu} - i(1 + C_{+})\ket{\omega_k^{(\downarrow, \uparrow)}}_2\ket{\nu}] \\ \nonumber
& + \frac{\beta_1\beta_2}{\sqrt{2}} \sum_{k,j,\mu,\nu}(\ket{\omega_k^{(\mu, \nu)}}_2\ket{\omega_j^{(\uparrow, \mu)}}_1 - i \ket{\omega_k^{(\mu, \nu)}}_2\ket{\omega_j^{(\downarrow, \mu)}}_1)\ket{\nu},
\end{align}
where $C_{-}=\sin{(\theta)}e^{-i\phi}+i \cos{(\theta)}$, $C_{+}=\sin{(\theta)}e^{i\phi}+i \cos{(\theta)}$, 
\begin{equation}
    \ket{\omega_j^{(\mu, \nu)}}_1 \coloneqq \int_0^{T_g} ds \lambda_j^{(\mu, \nu)}(T_g, s) b^\dagger_j(s) \ket{0},
\end{equation}
and
\begin{equation}
    \ket{\omega_j^{(\mu, \nu)}}_2 \coloneqq \int_{T_{g}}^{2T_g} ds \lambda_j^{(\mu, \nu)}(T_g, s-T_{g}) b^\dagger_j(s) \ket{0},
\end{equation}
with $\mu,\nu$ spanning the spin basis and $j,k$ the photon polarization basis. In the rest of this section, we will consider the wavepackets of both input photons being decreasing exponentials, $\xi(t) \coloneqq \sqrt{\Gamma}e^{\frac{-\Gamma t}{2}}$ with $T_{g}\gg \Gamma^{-1}$. 
\\
\\
To compare the ideal protocol with the realistic one, we compute the fidelity of $\ket{\Phi(2T_g)}$ with the final state of the ideal gate [Eq.~\eqref{ideal_cz}]:

\begin{align}\label{gate_fid}
\mathcal{F}_{\text{CZ}}(\alpha,\beta)=&\bigg| A|\alpha_1\alpha_2|^2 + B|\alpha_2\beta_1|^2 + C|\alpha_1\beta_2|^2 + D|\beta_1\beta_2|^2 \bigg|^2
\end{align}
with 
\begin{align}\label{eq_A-D}
A &= C_{-}; \\
B &= \frac{(-\Lambda^{(\uparrow, \uparrow)} + i\Lambda^{(\downarrow, \uparrow)})(1 - i\cos{(\theta)})}{\sqrt{2}}\nonumber \\ \nonumber &+\frac{ (\Lambda^{(\downarrow, \downarrow)}  + i \Lambda^{(\uparrow, \downarrow)})\sin{(\theta)}e^{-i\phi}}{\sqrt{2}};\\ \nonumber
C &= \frac{(C_{-}-1)\Lambda^{(\uparrow, \uparrow)} +i(1 + C_{+})\Lambda^{(\downarrow, \uparrow)}}{\sqrt{2}};\nonumber \\ \nonumber
D &= \Lambda^{(\uparrow, \downarrow)}\Lambda^{(\downarrow, \uparrow)} + \Lambda^{(\uparrow, \uparrow)}\Lambda^{(\uparrow, \uparrow)}- i \Lambda^{(\downarrow, \downarrow)}\Lambda^{(\downarrow, \uparrow)} - i \Lambda^{(\downarrow, \uparrow)}\Lambda^{(\uparrow, \uparrow)}; \nonumber\\
\text{and}\nonumber\\
&\Lambda^{(\mu, \nu)} \coloneqq \int_0^{T_g} ds \xi^*(s)\lambda_R^{(\mu, \nu)}(T_g, s)
\end{align}
being the overlap between the input and the output photons wavepackets.

As before, we eliminate the fidelity's dependence on the choice of the photons initial states, namely the parameters $\alpha\equiv(\alpha_1;\alpha_2)$ and $\beta\equiv(\beta_1,\beta_2)$, by integrating $\mathcal{F}_{\text{CZ}}(\alpha,\beta)$ over the uniform measures of their Bloch spheres:
\begin{equation}
    \begin{aligned}\label{eq_fidelity_gate_Bloch_average}
        \mathcal{F}_{\text{CZ}}= &\frac{1}{9}(\abs{A}^2 + \abs{B}^2 + \abs{C}^2 + \abs{D}^2) + \\ 
        & \frac{1}{18}(AB^* + AC^* + BD^* + CD^* + h.c) + \\
        & \frac{1}{36}(AD^* + BC^* + h.c).
    \end{aligned}
\end{equation}
The above expression shows that, as soon as $A=B=C=D=e^{i\delta}$, the fidelity goes to one. As shown in Appendix~\ref{App:C}, this condition is always verified in the ideal protocol: no Overhauser field, i.e. $\textbf{n}=\hat{x}$ and $\Omega_e = \Omega_g$; monochromatic photons, i.e. $\Gamma/\gamma\approx0$; instantaneous spontaneous emission, i.e. $\gamma/\Omega_{g}\approx0$; then summarizing, the realistic protocol converges to the ideal one when $\gamma \gg \Gamma \gg \Omega_g \approx \Omega_e$. 

\begin{figure}
    \centering
    \includegraphics[width=1\linewidth]{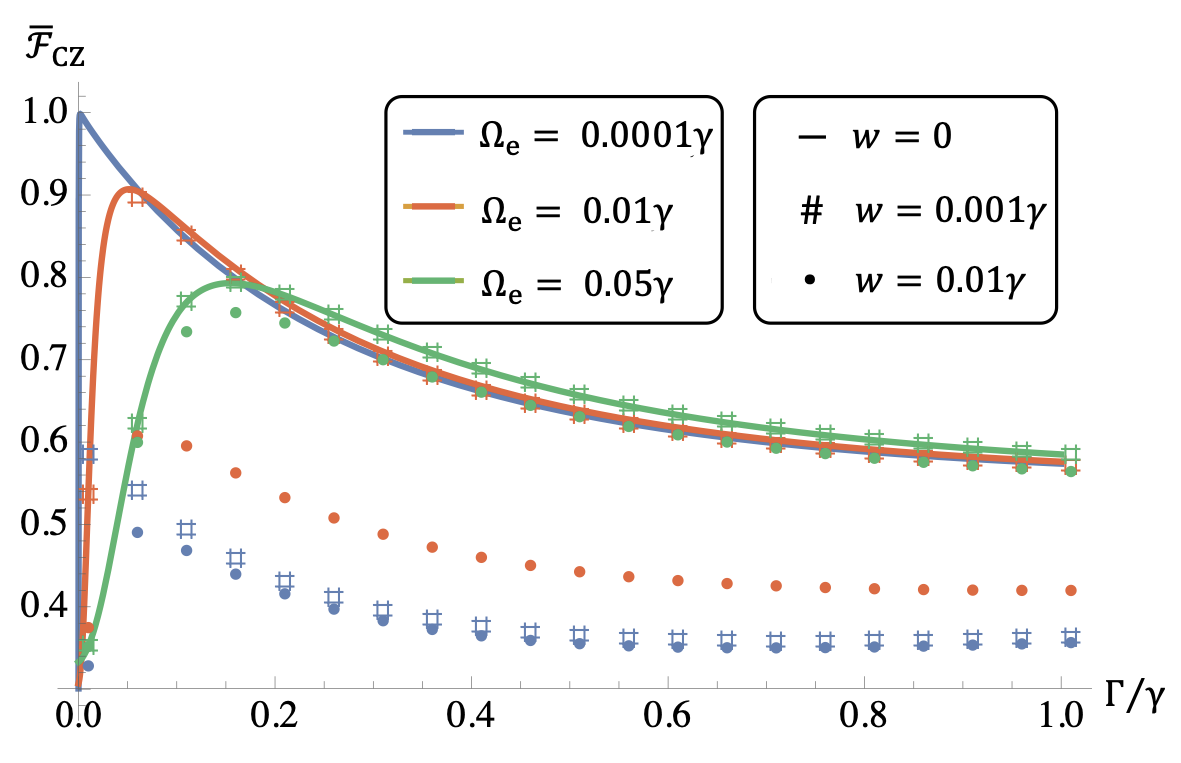}
    \caption{Fidelity upper bound of the CZ photon-photon gate varying the photons' bandwidth $\Gamma$ for different values of external magnetic field $\hbar\Omega_e/( g_{\text{tr}}\mu_B)= B^{\text{ext}}$, and spin coherence time $T_{2}^{*}=w^{-1}$. The curves are obtained from the numerical integration of Eq.~\eqref{eq_fidelity_gate_Bloch_average} with Overhauser field distribution centered around $\bar{\Omega}_g=\Omega_e$ (see the text).}
    \label{figIII}
\end{figure}

Aside from the ideal limit, we account for the Overhauser field's randomness, as before, via statistical average with the Gaussian distribution of width $w$. Due to the external magnetic field, this time the center of the Gaussian distribution of the electron spin precession frequencies is different from zero, we denote it with $\bar{\Omega}_{g}$. The fidelity is plotted in Fig.~(\ref{figIII}) as a function of the bandwidth $\Gamma/\gamma$ for different values of $\Omega_e=\bar{\Omega}_{g}$. The plot shows that, as expected from the chain of inequalities $\gamma \gg \Gamma \gg \Omega_e $, by increasing the magnitude of the external magnetic field  $\hbar \Omega_e/( g_{\text{tr}}\mu_B)= B^{\text{ext}}$, the maximum of the fidelity moves towards greater values of the photons' bandwidth $\Gamma/\gamma$. Notice that such dependence on the incoming photons bandwidth can only be captured by a model that goes beyond the monochromatic approximation. 

The plot also shows that when the Overhauser field is taken into account, an optimal performance requires $\Omega_e=\bar{\Omega}_g \gg w$, this reflects the fact that the detrimental impact of the Overhauser field becomes negligible in a strong external field. In a realistic SPI, where the spin decoherence rate $w$ is fixed by the QD structure, this optimization can only be achieved by increasing the external magnetic field. At the same time, however, $\Omega_e$ must remain smaller than the photon bandwidth $\Gamma$ that in turn must be smaller than the spontaneous emission rate $\gamma$, making the whole optimization very challenging. As shown in the plot, in order to obtain a fidelity close to $90\%$, in the absence of Overhauser field, one needs to work with a spin precession frequency at least two orders of magnitude smaller than the spontaneous emission rate (blue and orange lines in figure). However, with such a weak external field, the presence of even a very small Overhauser field is highly detrimental for the gate performance. As mentioned in the previous section, a realistic value of $w/\gamma$ is $\sim 0.1$, considering $w^{-1}\sim 1$ ns, and $\gamma^{-1}\sim 100$ ps. Yet, our results show that, even in the scenario of a SPI having a spin decoherence rate 100 times smaller than the spontaneous emission rate ($w/\gamma=0.01$), the fidelity upper bound barely reaches $0.8$ (green line) which is far below the regime where even near-term algorithms can produce useful output. 

\subsection{Lindner-Rudolph protocol for cluster states generation}\label{LR}

\begin{figure}
    \centering
    \includegraphics[width=1\linewidth]{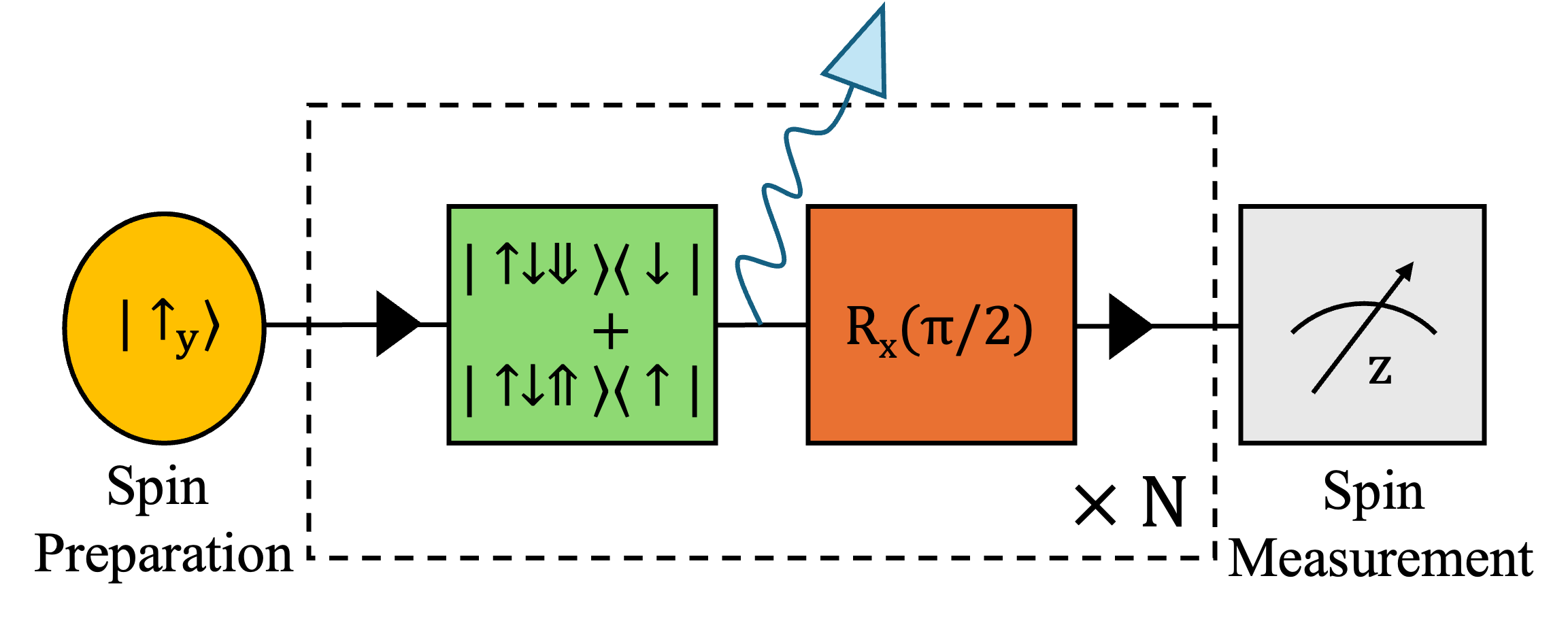}
    \caption{Schematic representation of the ideal Lindner-Rudolph protocol. The electron spin is prepared in $\ket{\uparrow_y}= \left(\ket{\uparrow} +i \ket{\downarrow}\right)/\sqrt{2}$. Then a "unit step" is performed $N$ times to generate an entangled state of $N$ photons plus spin. The unit step consists of a classical excitation of the QD, followed by spontaneous emission, followed by a $\pi/2$ rotation of the electron spin around the $x$ axis. The joint state of the spin and the $N$ photons is given by Eq.~\eqref{ideal_LR}. After this the electron spin is measured and disentangled from the $N$ photons resulting in an $N$ photon cluster state.}
    \label{ClusterSchematic}
\end{figure}
\subsubsection*{Ideal protocol} 

The Lindner-Rudolph (LR) protocol generates linear clusters of propagating photons entangled in polarization, having, in principle, arbitrary length~\cite{lindner_proposal_2009}. Each photon of the cluster is generated by an instantaneous excitation of the QD's energy levels followed by spontaneous emission and spin precession. A cluster of N photons is then generated by N repetitions of this unit step as shown in Fig.~(\ref{ClusterSchematic}). In the ideal protocol, the sequence starts with the SPI prepared in the trion spin superposition state, $\ket{\Psi(0)}=\ket{e}\ket{\uparrow_y}$, where $\ket{\uparrow_y}=(\ket{\uparrow}+i\ket{\downarrow})/\sqrt{2}$. Then the spontaneous emission, considered as instantaneous with respect to the electron spin precession, creates a maximally entangled spin-photon state, $\ket{\Psi(0^{+})}=\ket{g}\left(\ket{\uparrow,1_{R}}+i\ket{\downarrow,1_{L}}\right)/\sqrt{2}$. In the ideal protocol, there is no Overhauser field and electron and trion have equal Land\'e factors, i.e. $\Omega_g=\Omega_e=\Omega$ and $\textbf{n}=\lbrace 1,0,0\rbrace$, hence the action of $H_{\text{s}}$ for a time $T_{1}=\pi/(2 \Omega)$ implements a rotation on the spin qubit of an angle $\pi/2$ around the $x$ axis. The SPI wavefunction at time $t=T_1$ is then 
\begin{align}\label{eq_wf1_ideal}
\ket{\Psi_1}=\frac{1}{\sqrt{2}}\ket{g}\left( \ket{\uparrow_y}\ket{1_L}_1 +\ket{\downarrow_y}\ket{1_R}_1\right),
\end{align}
 where $ \ket{1_R(L)}_1=\sqrt{\gamma} \int_{0}^{T_1} dt  e^{-\gamma t/2 }b^{\dagger}_{R(L)}(t)\ket{0}$ is a single-photon state normalized in $[0,T_1]$, condition guaranteed by the assumption of instantaneous emission i.e. $\Omega/\gamma\approx0$ ($e^{-\gamma T_1}\approx 0$). Then a classical pulse re-excites the QD such that a second photon is emitted followed by another spin precession for time interval equal to $T_1$. The state at $t=2T_{1}$ is then 
\begin{equation}\label{eq_wf2_ideal}
\begin{aligned}
    \ket{\Psi_2}=\frac{1}{2}\ket{g}&\left(\ket{\downarrow_y}\ket{1_R}_1\ket{1_R}_2 + \ket{\downarrow_y}\ket{1_L}_1\ket{1_R}_2 \right.\\ &- \left.\ket{\uparrow_y}\ket{1_R}_1\ket{L_R}_2 + \ket{\uparrow_y}\ket{1_L}_1\ket{1_L}_2\right),
\end{aligned}
\end{equation}
where the subscript 2 means that the photon's state has support in $[T_1,2T_1]$. The wavefunction of the joint system after N repetitions of the ideal unit step can be written in the compact matrix product form:
\begin{align}\label{ideal_LR}
\ket{\Psi_N}= \ket{g}\begin{pmatrix} \frac{1}{\sqrt{2}}& \frac{i}{\sqrt{2}}\end{pmatrix} \cdot \left( \mathcal{W}_1 \cdot \mathcal{W}_2 \cdot ...\cdot  \mathcal{W}_N\right) \cdot \begin{pmatrix} \ket{\uparrow}\\\ket{\downarrow}\end{pmatrix} \\
\equiv \ket{g}\begin{pmatrix} \frac{1}{\sqrt{2}}& \frac{i}{\sqrt{2}}\end{pmatrix} \prod_{j=1}^N(\mathcal{W}_j)\begin{pmatrix} \ket{\uparrow}\\\ket{\downarrow}\end{pmatrix} \nonumber
\end{align}
with
\begin{align}
\mathcal{W}_m=\frac{1}{\sqrt{2}}\begin{pmatrix}
\ket{1_{R}}_m & -i \ket{1_R}_m \\
-i\ket{1_L}_m & \ket{1_L}_m 
\end{pmatrix}
\end{align}
\color{black}
The row vector in the matrix product is the initial state of the spin, in this case $\ket{\uparrow_y}$, with $\ket{\uparrow}=\begin{pmatrix} 1 & 0\end{pmatrix}, \ket{\downarrow}=\begin{pmatrix} 0 & 1\end{pmatrix}$.
\\
\\
In the Lindner-Rudolph seminal paper, the authors consider the deviation from the ideal protocol due to a finite ratio $\Omega/\gamma$. As mentioned in Sec.~\ref{secII}, this deviation can be linked to the WAY theorem~\cite{Loveridge2011Measurement}. The latter says that it is impossible to obtain a perfect one-to-one mapping between two interacting quantum systems when their interaction Hamiltonian does not commute with their full Hamiltonian: as $H_{\text{s}}$ does not commute with $V$, the mapping $\ket{\uparrow(\downarrow)}\rightarrow\ket{1_{R(L)}}$ cannot be perfect. Lindner and Rudolph showed that, for $\Omega_e=\Omega_g=\Omega$ and $w/\gamma\approx 0$, the sole effect of the finite ratio $\Omega/\gamma \neq 0$ induces a $X$ Pauli error on the spin qubit. In the next section we show that our Hamiltonian approach gives the same result. As for the spin decoherence induced by the Overhauser field, they model it as a further $X$ Pauli error on the spin qubit whose probability depends solely on $w/\gamma$. In their top-down approach, the overall error probability of the protocol's unit step is the product of the probabilities of the two errors considered as independent. As we show in the next section, our microscopic model allows us to go beyond this approximation, and it also accounts naturally for the loss of fidelity due to deformations of the photonic wavepackets. 

\subsubsection*{Hamiltonian model}

Throughout this section, we will assume that the classical pulse exciting the QD is linearly polarized and very short with respect to the trion's lifetime, hence producing a nearly instantaneous population swap from electron to trion for both spin projections. Furthermore, we consider the time $T_1=\pi/(2\Omega_{g})$ of each electron spin precession to be longer than the trion's lifetime, yet keeping their ratio finite. We can then write the SPI wavefunction after N repetitions of the realistic unit step as:
\begin{align}\label{mp_real}
\ket{\Psi(N T_1)}=\ket{g}\begin{pmatrix} \frac{1}{\sqrt{2}}&\frac{i}{\sqrt{2}}\end{pmatrix} \prod_{m=1}^N (W_m) \begin{pmatrix} \ket{\uparrow}\\\ket{\downarrow}\end{pmatrix} 
\end{align}
with
\begin{align}
W_m=\begin{pmatrix}
\sum_{i=R,L}\ket{\Phi_{i}^{(\uparrow,\uparrow)}}_m & \sum_{i=R,L}\ket{\Phi_{i}^{(\uparrow,\downarrow)}}_m \\
\sum_{i=R,L}\ket{\Phi_{i}^{(\downarrow,\uparrow)}}_m & \sum_{i=R,L}\ket{\Phi_{i}^{(\downarrow,\downarrow)}}_m, 
\end{pmatrix}
\end{align}
where we defined the un-normalized field's wavefunctions as:
\begin{align}
\ket{\Phi^{(\zeta,\mu)}_{i}}_m=\int_{(m-1)T_1}^{m T_1} dt f_{i}^{(\zeta,\mu)}(m T_1,t+T_1-mT_1)b^{\dagger}_i(t)\ket{0};
\end{align}
the subscript $m$ refers to the fact that the photon is emitted in the interval $[(m-1)T_1,m T_1]$, and the coefficients $f_{i}^{(\zeta,\mu)}(t,t')$ are given in Appendix~\ref{App:A}.
Notice that, in the ideal protocol, the field's states $\ket{\Phi^{(\uparrow, \mu)}_{L}}$ and $\ket{\Phi^{(\downarrow,\zeta)}_{R}}$ are never populated for any $\mu,\zeta=\uparrow, \downarrow$. The probability to find a photon in these states corresponds to the conditional probabilities of detecting an L (R) photon after preparing the spin in $\ket{\uparrow}$ ($\ket{\downarrow}$), namely $P(\downarrow|R)=P(\uparrow|L)$ where $P(\zeta|j)=\sum_{\mu} \langle \Phi^{(\zeta, \mu)}_{j}\vert \Phi^{(\zeta, \mu)}_{j}\rangle$, it reads
\begin{align}\label{eq_Perror}
P(\downarrow|R)=P(\uparrow|L)=\frac{(\Omega_e/\gamma)^2}{2\left[1+(\Omega_e/\gamma)^2\right]}.
\end{align}
The above expression is the probability derived by Lindner and Rudolph in their seminal paper for the bit-flip (X Pauli) error as mentioned in the previous subsection. 
\\
\\
Using the matrix product form of the ideal and the real state, it is possible to find a compact expression for the fidelity's upper bound after $N$ iterations of the protocol. Assuming that the final spin measurement is perfectly projective,  the fidelity is given by the modulus square of the states overlap, that can be written as:
\begin{align}\label{eq_overlapN}
\langle \Psi(N T_1) \ket{\Psi_N}=\begin{pmatrix} \frac{1}{\sqrt{2}}&\frac{i}{\sqrt{2}}\end{pmatrix} \left(\Pi_{m=1}^{N} \mathcal{W}_m \right)\left(\Pi_{m=1}^{N} W_m \right)^\dagger\begin{pmatrix} \frac{1}{\sqrt{2}}\\\frac{-i}{\sqrt{2}}\end{pmatrix} 
\end{align}
where again the first and the last vectors correspond to the state of the spin at $t = 0$.

The above expression shows that for any number of iterations the fidelity can be derived by combining suitably the 8 overlaps :
\begin{align}
O^{(\zeta,\mu)}_{j}=\left\langle \textbf{1}_j \vert\Phi^{(\zeta,\mu)}_{j}\right\rangle=\int_{0}^{T_1} dt \sqrt{\gamma} e^{-\gamma t/2} f_{j}^{(\zeta,\mu)*}(T_1,t).
\end{align}
For instance, using Eq.~\eqref{eq_overlapN} is easy to find that the fidelity at time $T_1$: 
\begin{equation}\label{eq_FLR1}
\begin{aligned}
    \mathcal{F}_{LR}^{(1)}&=|\langle \Psi_1 \vert \Psi(T_1) \rangle|^2\\ &=\frac{1}{2\sqrt{2}}|O_R^{(\uparrow,\uparrow)} +  O_L^{(\uparrow,\uparrow)} + i O_R^{(\uparrow,\downarrow)} - i O_L^{(\uparrow,\downarrow)} \\&+ i O_R^{(\downarrow,\uparrow)} + i O_L^{(\downarrow,\uparrow)} - O_R^{(\downarrow,\downarrow)} + O_L^{(\downarrow,\downarrow)}|^2
\end{aligned}
\end{equation}
Similarly the fidelity after 2 iterations is:
\begin{equation}\label{eq:FLR2}
\begin{aligned}
    &\mathcal{F}_{LR}^{(2)}=|\langle \Psi_2\vert \Psi(2T_1)\rangle |^2\\ 
&=\vert \frac{1}{4} \sum_{\mu=\uparrow,\downarrow} [\left(O_R^{(\uparrow,\mu)} + i O_R^{(\downarrow, \mu)}\right)\left( O_R^{(\mu,\uparrow)} - O_L^{(\mu,\uparrow)} + i O_R^{(\mu,\downarrow)} +i O_L^{(\mu,\downarrow)}\right) \\& +\left(O_L^{(\uparrow, \mu)} + i O_L^{(\downarrow, \mu)}\right) \left(O_R^{(\mu,\uparrow)} + O_L^{(\mu,\uparrow)} + i O_R^{(\mu,\downarrow)} -i O_L^{(\mu,\downarrow)}\right)]\vert^2
\end{aligned}
\end{equation}

In the absence of Overhauser noise, setting $\textbf{n}=\hat{x}$ and $\Omega_g= k \Omega_e$ with $k=g_{\text{el}}/g_{\text{tr}}$ being the ratio between electron and trion Land\'e factors, it is possible to find a compact expression for the fidelity after the $j$-th step (at time $t=jT_1$) that we denote $\mathcal{F}_{\text{LR, id}}^{(j)}$:

\begin{equation}
    \mathcal{F}_{\text{LR, id}}^{(j)}= \mathcal{F}_{\text{LR, id}}^{(1)} f^{j-1} + O(\Omega_e^4/\gamma^4)
\end{equation}
with
\begin{equation}
\begin{aligned}
        \mathcal{F}_{\text{LR, id}}^{(1)}&
        &= \left[1 - \left(\frac{1+k^2}{2}\right)P(\downarrow\vert R)\right]^2 + O(\Omega_e^4/\gamma^4),
\end{aligned}
\end{equation}
 and 
\begin{equation}
    f\coloneqq\frac{\mathcal{F}_{\text{LR, id}}^{(2)}}{\mathcal{F}_{\text{LR, id}}^{(1)}} = \left[1 -\frac{(1 - k + k^2)}{2}P(\downarrow\vert R)\right]^2 + O(\Omega_e^4/\gamma^4)
\end{equation}
The above expression, for $k=1$, gives the fidelity obtained in the seminal paper using the bit-flip error. Setting experimentally realistic values, such as $k=2$ and $\Omega_e/\gamma =0.1$, it gives a fidelity above $75\%$ for up to 15 photons.
\begin{figure}
    \centering
    \includegraphics[width=1\linewidth]{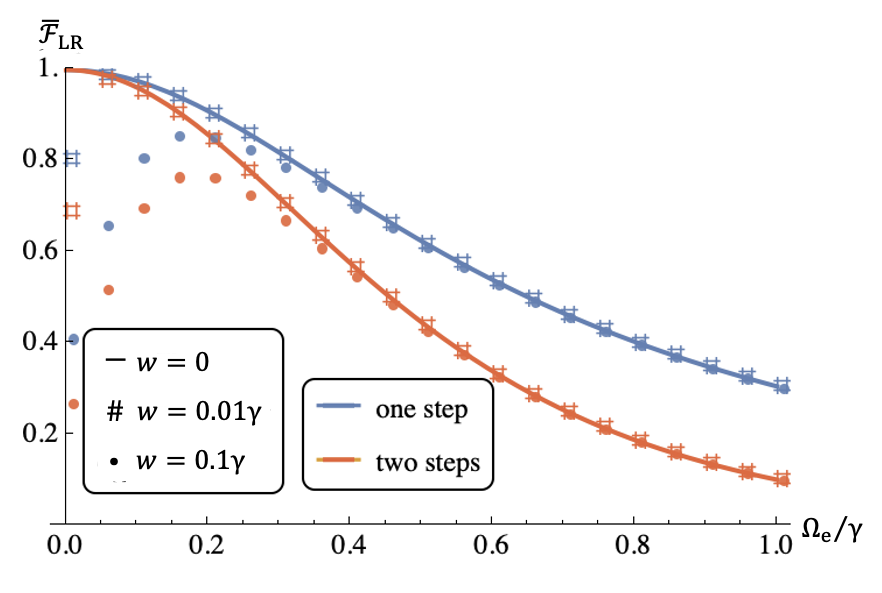}
    \caption{Fidelity upper bound of the final states of LR protocol after one (blue) and two (orange) steps. The curves are obtained respectively from the numerical integration of Eq.~\eqref{eq_FLR1} and Eq.~\eqref{eq:FLR2} with Overhauser field distribution of width $w$ centered around $\bar{\Omega}_g=2\Omega_e$ varying $\Omega_e$.} 
    \label{fig4}
\end{figure}
\\
\\
As in the previous sections, we finally account for the presence of the Overhauser field by taking the statistical average of the fidelities [Eq.~\eqref{eq_FLR1} and Eq.~\eqref{eq:FLR2}] with the isotropic Gaussian distribution of width $w$. The curves, obtained via numerical integration, are shown in Fig.~(\ref{fig4}). The plots show clearly that the Overhauser field affects the fidelity significantly only when $w$ becomes comparable with $\Omega_{e}$ i.e. when the Overhauser field cannot be treated as a small perturbation of the external field. Thus the optimal value of the tunable parameter $\Omega_e$ depends on the spin decoherence rate $w$ of the device. 

Let us notice that, at the working point of Ref.~\cite{Fioretto2022}, $k=2$, $\Omega_e/\gamma \approx 0.2$ and $w/\gamma \approx 0.1$, our curves (dotted data sets) predict $\overline{\mathcal{F}^{(1)}}_{\text{LR}}=0.85$ and $\overline{\mathcal{F}^{(2)}}_{\text{LR}}=0.76$. These values, as expected, are slightly bigger than those reported in the experimental paper, resp. $0.8$ and $0.63$. We see that for $w/\gamma=0.01$ we can tune the magnetic field to obtain fidelity upper bounds between $0.9-0.99$ for $N=1,2$. Using the slight underestimate of $p=0.01$ error per unit step, such cluster states are sufficient for fault tolerant quantum computation \cite{ibm_cluster}. 

\section{Conclusion}

We solved the Hamiltonian dynamics of a QD SPI interacting with photonic wavepackets in the presence of magnetic fields. We model the solid-state SPI as a degenerate four level system in the absence of external fields where the ground states correspond to orthogonal projections of an electron spin and the excited states to those of a trion spin. Our approach considers both an external tunable magnetic field, and an Overhauser field, namely the effective magnetic field modeling the electron's hyperfine interaction with the nuclei of the atoms forming the QD. Following the seminal Merkulov-Efros-Rosen semiclassical model, the latter is treated as a static isotropic white noise, and we account for its randomness by taking statistical averages of the quantum mechanical observables over all its possible configurations, with the spin decoherence rate then given by the width of the white noise distribution. Using this model, we analytically evaluated the performance of three key technological tasks achievable with the SPI: generation of coherent superpositions of 0- and 1-photon states, photon-photon controlled Z gate, and the Lindner-Rudolph protocol to generate linear photonic clusters. Our results show that the spin decoherence due to the Overhauser field does not impact substantially the fidelity of superpositions of 0- and 1-photon states. In contrast, the photon-photon gate is found to be extremely sensitive to it: the gate functions reliably only when the Overhauser field is absent, and is very challenging to optimize even when its intensity is one order of magnitude lower than in state-of-the-art devices. This is due to the fact that, on the one hand, noiseless photon-scattering (without wavepackets deformations) requires very long photons, and on the other hand, mitigation of the spin decoherence requires very short scattering dynamics. Finally, we show that the Lindner–Rudolph protocol for cluster-state generation is comparatively robust against spin decoherence. Recently Ref.~\cite{ibm_cluster} proposed a protocol of measurement-based quantum computation~\cite{Briegel2009} that is fault tolerant with very high thresholds. The resource states used are one dimensional cluster states. Using the error analysis of the seminal paper~\cite{lindner_proposal_2009}, it is found in simulations that the LR protocol is sufficiently robust to generate the required quality of cluster states for fault tolerant quantum computation. The present analysis can be used to refine such simulations by using more accurate resource states predicted by our model. Our results provide theoretical fidelity bounds for passive implementations indicating how active error-mitigation strategies could further enhance protocols quality if required. On the one hand, the present multi mode analysis of propagating electromagnetic fields paves the way for optimization strategies based on active pulse-shaping~\cite{Scarani2011}: especially for the CZ gate, identifying and engineering the optimal wavepackets shapes for the input photons would arguably improve the protocol performance representing a promising outlook. On the other hand, our results show that the considered protocols work better when the Overhauser field is a small perturbation of the external magnetic field, which in turn must be weak to make the spin precession slower than the spontaneous emission. This creates a highly constrained parameter space where high-precision techniques for in-situ control of the magnetic field~\cite{Pang2025} could be a viable pathway towards better performances. We conclude by noting that, even though the analysis was tailored for negatively charged QD SPI, it holds for all systems with similar structure of the energy levels and selection rules. From a methodological point of view, our results prove that adopting a Hamiltonian analysis is not only feasible, but fruitful for accurate benchmarking of protocols for computation with propagating photons. 
\\
\\
\textit{Acknowledgments --} A.A., HKN, and TA acknowledge the National Research Foundation, Singapore through the National Quantum Office, hosted in A*STAR, under its Centre for Quantum Technologies Funding Initiative (S24Q2d0009). A.A. acknowledges the Plan France 2030 through the projects NISQ2LSQ (Grant ANR-22-PETQ-0006) and OQuLus (Grant ANR-22-PETQ-0013). TA and HKN are also supported in part by the NRF-ANR joint QuRes project (NRF2021-NRF-ANR005).

%

\begin{widetext}
\appendix
\section{Explicit expression of the spin-photon wavefunction}\label{App:A}
First we give the expressions for spontaneous emission. Note that when there are two time arguments, the first argument is the observation time and the second argument is the time mode described by the wave function:
\begin{align}
f_0^{(\zeta, \mu)} = &e^{-\frac{\gamma t}{2}}\left( \cos{\left(\frac{\Omega_e t}{2}\right)}\delta_{\mu, \zeta} - i \sin\left(\frac{\Omega_et}{2}\right)(1 - \delta_{\mu, \zeta})\right)\nonumber\\
f_{1_{R}}^{(\uparrow,\uparrow)}(t,t^{\prime})= \sqrt{\gamma}& e^{-\frac{1}{2}(\gamma t^{\prime})}\cos\left(\frac{\Omega_{e}t^{\prime}}{2}\right)\left(\cos\left(\frac{1}{2}\Omega_{g}(t-t^{\prime})\right)-in_{z}\sin\left(\frac{1}{2}\Omega_{g}(t-t^{\prime})\right)\right)=\left(f_{1_{L}}^{(\downarrow,\downarrow)}(t,t^{\prime})\right)^{*}\nonumber\\
f_{1_{R}}^{(\uparrow,\downarrow)}(t,t^{\prime})= & -i\sqrt{\gamma}(n_{x}+in_{y})e^{-\frac{1}{2}(\gamma t^{\prime})}\cos\left(\frac{t^{\prime}\Omega_{e}}{2}\right) \sin\left(\frac{1}{2}\Omega_{g}(t-t^{\prime})\right)=-\left(f_{1_{L}}^{(\downarrow,\uparrow)}(t,t^{\prime})\right)^{*}\nonumber\\
f_{1_{L}}^{(\uparrow,\uparrow)}(t,t^{\prime})= &i\sqrt{\gamma}(n_{y}+in_{x})e^{-\frac{1}{2}(\gamma t^{\prime})}\sin\left(\frac{t^{\prime}\Omega_e}{2}\right)\sin\left(\frac{1}{2}\Omega_{g}(t-t^{\prime})\right)=\left(f_{1_{R}}^{(\downarrow,\downarrow)}(t,t^{\prime})\right)^{*}\nonumber\\
f_{1_{L}}^{(\uparrow,\downarrow)}(t,t^{\prime})= & -i\sqrt{\gamma}e^{-\frac{1}{2}(\gamma t^{\prime})}\sin\left(\frac{t^{\prime}\Omega_e}{2}\right)\left(\cos\left(\frac{1}{2}\Omega_{g}(t-t^{\prime})\right)+in_{z}\sin\left(\frac{1}{2}\Omega_{g}(t-t^{\prime})\right)\right)=-\left(f_{1_{R}}^{(\downarrow,\uparrow)}(t,t^{\prime})\right)^{*}\nonumber
\end{align}
The overlaps $O_j^{(\zeta, \mu)}$ are simply time integrals of these multiplied by $\sqrt{\gamma}e^{-\gamma t'/2}$ from $0$ to $t \coloneqq T_1 = \pi/(2\Omega_g^{(0)})$. Now we list the expressions for single-photon scattering. 
\begin{align}
    \lambda_R^{(\uparrow, \uparrow)}(t, t^\prime) = &\xi(t^\prime)\left[\cos\left(\frac{\Omega_g t}{2}\right) - i n_z \sin\left(\frac{\Omega_g t}{2}\right)\right] \nonumber \\ 
    - &\gamma \left[ \cos \left( \frac{\Omega_g(t - t^\prime)}{2}\right) - i n_z \sin \left( \frac{\Omega_g(t - t^\prime)}{2}\right)\right] \int_0^{t^\prime} dt^{\prime\prime} \xi(t^{\prime \prime})e^{-\frac{\gamma(t^\prime - t^{\prime \prime})}{2}} \cos \left( \frac{\Omega_e (t^\prime - t^{\prime \prime})}{2}\right)\left[\cos \left( \frac{\Omega_g t^{\prime\prime}}{2} \right) - i n_z \sin \left( \frac{\Omega_g t^{\prime\prime}}{2} \right)\right] \nonumber \\
    \lambda_R^{(\downarrow, \uparrow)}(t, t^\prime) = &-i\xi(t^\prime)\left[n_x - in_y\right]\sin \left(\frac{\Omega_g t}{2}\right) \nonumber \\ 
    +i &\gamma \left[ \cos \left( \frac{\Omega_g(t - t^\prime)}{2}\right) - i n_z \sin \left( \frac{\Omega_g(t - t^\prime)}{2}\right)\right] \int_0^{t^\prime} dt^{\prime\prime} \xi(t^{\prime \prime})e^{-\frac{\gamma(t^\prime - t^{\prime \prime})}{2}} \cos \left( \frac{\Omega_e (t^\prime - t^{\prime \prime})}{2}\right)\left[n_x - i n_y\right]\sin\left(\frac{\Omega_g t^{\prime\prime}}{2}\right) \nonumber \\ 
    \lambda_R^{(\uparrow, \downarrow)}(t, t^\prime) = &-i \xi(t^\prime)\left[n_x + i n_y\right]\sin\left(\frac{\Omega_g t}{2}\right) \nonumber \\ 
    +i &\gamma \left[ n_x + i n_y \right] \sin \left( \frac{\Omega_g(t - t^\prime)}{2}\right) \int_0^{t^\prime} dt^{\prime\prime} \xi(t^{\prime \prime})e^{-\frac{\gamma(t^\prime - t^{\prime \prime})}{2}} \cos \left( \frac{\Omega_e (t^\prime - t^{\prime \prime})}{2}\right)\left[\cos \left( \frac{\Omega_g t^{\prime\prime}}{2} \right) - i n_z \sin \left( \frac{\Omega_g t^{\prime\prime}}{2} \right)\right] \nonumber \\ 
    \lambda_R^{(\downarrow, \downarrow)}(t, t^\prime) = &\xi(t^\prime)\left[\cos\left(\frac{\Omega_g t}{2}\right) - i n_z \sin\left(\frac{\Omega_g t}{2}\right)\right] \nonumber \\ 
    +i &\gamma \left[ n_x^2 +  n_y^2 \right] \sin \left( \frac{\Omega_g(t - t^\prime)}{2}\right) \int_0^{t^\prime} dt^{\prime\prime} \xi(t^{\prime \prime})e^{-\frac{\gamma(t^\prime - t^{\prime \prime})}{2}} \cos \left( \frac{\Omega_e (t^\prime - t^{\prime \prime})}{2}\right)\sin \left( \frac{\Omega_g t^{\prime\prime}}{2} \right) \nonumber \\ 
    \lambda_L^{(\uparrow, \uparrow)}(t, t^\prime) = 
    &\gamma \left[ n_x - i n_y \right] \sin \left( \frac{\Omega_g(t - t^\prime)}{2}\right) \int_0^{t^\prime} dt^{\prime\prime} \xi(t^{\prime \prime})e^{-\frac{\gamma(t^\prime - t^{\prime \prime})}{2}} \sin \left( \frac{\Omega_e (t^\prime - t^{\prime \prime})}{2}\right)\left[\cos \left( \frac{\Omega_g t^{\prime\prime}}{2} \right) - i n_z \sin \left( \frac{\Omega_g t^{\prime\prime}}{2} \right)\right] \nonumber \\ 
    \lambda_L^{(\downarrow, \uparrow)}(t, t^\prime) = 
    &-i\gamma \left[ n_x - i n_y \right]^2 \sin \left( \frac{\Omega_g(t - t^\prime)}{2}\right) \int_0^{t^\prime} dt^{\prime\prime} \xi(t^{\prime \prime})e^{-\frac{\gamma(t^\prime - t^{\prime \prime})}{2}} \sin \left( \frac{\Omega_e (t^\prime - t^{\prime \prime})}{2}\right)\sin \left( \frac{\Omega_g t^{\prime\prime}}{2} \right) \nonumber \\ 
    \lambda_L^{(\uparrow, \downarrow)}(t, t^\prime) = 
    &i\gamma \left[ \cos \left( \frac{\Omega_g(t - t^\prime)}{2}\right) + i n_z \sin \left( \frac{\Omega_g(t - t^\prime)}{2}\right)\right] \int_0^{t^\prime} dt^{\prime\prime} \xi(t^{\prime \prime})e^{-\frac{\gamma(t^\prime - t^{\prime \prime})}{2}} \sin \left( \frac{\Omega_e (t^\prime - t^{\prime \prime})}{2}\right)\left[\cos \left( \frac{\Omega_g t^{\prime\prime}}{2} \right) - i n_z \sin \left( \frac{\Omega_g t^{\prime\prime}}{2} \right)\right] \nonumber \\
    \lambda_L^{(\downarrow, \downarrow)}(t, t^\prime) = 
    &\gamma (n_x - in_y)\left[ \cos \left( \frac{\Omega_g(t - t^\prime)}{2}\right) + i n_z \sin \left( \frac{\Omega_g(t - t^\prime)}{2}\right)\right] \int_0^{t^\prime} dt^{\prime\prime} \xi(t^{\prime \prime})e^{-\frac{\gamma(t^\prime - t^{\prime \prime})}{2}} \sin \left( \frac{\Omega_e (t^\prime - t^{\prime \prime})}{2}\right)\sin \left( \frac{\Omega_g t^{\prime\prime}}{2} \right) \nonumber
\end{align}

\section{Explicit derivation of the fidelity of the coherent single photon source}\label{App:B}

Recall that $\rho_\infty = \text{Tr}_{\text{spin}}[\ket{\Psi_\infty^{(\uparrow)}}\bra{\Psi_\infty^{(\uparrow)}}]$ and $\mathcal{F}_\text{PNS}(\alpha, \beta) = \bra{\Psi_\text{ideal}} \rho_\infty \ket{\Psi_\text{ideal}}$. This can be neatly written by instead equipping the ideal state artificially with spin components and simply computing the fidelity of this state with $\ket{\Psi_\infty^{(\uparrow)}}\bra{\Psi_\infty^{(\uparrow)}}$. We obtain 
\begin{align}\label{eq_Fid_photon_source}
        \mathcal{F}_{\text{PNS}}(\alpha,\beta)&=\sum_{\mu=\uparrow,\downarrow}|\alpha^{*}\bra{1_{R},\mu,g}\ket{\Psi^{(\uparrow)}_\infty})+\beta^{*}\bra{0,\mu,g} \ket{\Psi^{(\uparrow)}_{\infty}}|^2\\ \nonumber
        &=\sum_{\mu=\uparrow,\downarrow}|\alpha^{*}\bra{1_{L},\mu,g}\ket{\Psi^{(\downarrow)}_{\infty}}+\beta^{*}\bra{0,\mu,g} \ket{\Psi^{(\downarrow)}_{\infty}}|^2
    \end{align}
 where the second equality derives from the property
$f_{1_{R}}^{(\uparrow,\uparrow)}=\left(f_{1_{L}}^{(\downarrow,\downarrow)}\right)^{*}$ and $f_{1_{R}}^{(\uparrow,\downarrow)}=-\left(f_{1_{L}}^{(\downarrow,\uparrow)}\right)^{*}$ (see Appendix~\ref{App:A}). 

Let us first write the first member of Eq.~\eqref{eq_Fid_photon_source} as: 
\begin{align}\label{step1}
F=|\alpha|^4 A +|\alpha|^2 |\beta|^2 B + |\beta|^4 
\end{align}
with
\begin{align}
    &A = \abs{\bra{1_R}\ket{\phi_{R}^{(\uparrow, \uparrow)}}}^ 2 + \abs{\bra{1_R}\ket{\phi_{R}^{(\uparrow, \downarrow)}}}^ 2 \\
    &B = (\cos{\left(\frac{\Omega t_\infty}{2}\right)} + i \cos{(\theta)}\sin{\left(\frac{\Omega t_\infty}{2}\right)})\bra{1_R}\ket{\phi_{R}^{(\uparrow, \uparrow)}} + \nonumber \\
    & (i \cos{(\phi)}\sin{(\theta)} + \sin{(\phi)}\sin{(\theta)})\sin{\left(\frac{\Omega t_\infty}{2}\right)}\bra{1_R} \ket{\phi_{R}^{(\uparrow, \downarrow)}} + h.c.
\end{align}
Naively it looks like there is still a $t_\infty$ dependence, but this is not the case. We see that by writing explicitly the quantities $\bra{1_R}\ket{\phi_R^{\uparrow, \uparrow}}$ and $\bra{1_R}\ket{\phi_R^{\uparrow, \downarrow}}$ as a function of the two adimensional variable $x \coloneqq \frac{\Omega}{2 \gamma}$ and $b \coloneqq \gamma t_\infty$.
\begin{equation}
    \bra{1_R}\ket{\phi_R^{\uparrow, \uparrow}} = \frac{\cos(xb)[1 + ix \cos(\theta)] + \sin(xb)[x - i \cos(\theta)]}{1 + x^2}
\end{equation}
\begin{equation}
    \bra{1_R}\ket{\phi_R^{\uparrow, \downarrow}} = [i \sin(\theta)\cos(\phi) - \sin(\theta)\sin(\phi)]\frac{x \cos(xb) - \sin(xb)}{1 + x^2}
\end{equation}
Entering the above expressions into $A$ and $B$ we get:
\begin{align}
A=\frac{1}{1+x^2};~B=\frac{2}{1+x^2};
\end{align}
plugging those expressions for $A$ and $B$ into Eq.~\eqref{step1}, we obtain Eq.~\eqref{pns_fid}.

\section{Explicit derivation of ideal working point of CZ gate}\label{App:C}

Here we show into details how the chain of inequalities required for the optimization of the CZ gate, i.e. $\gamma \gg \Gamma \gg \Omega_g \approx \Omega_e$, emerges from the explicit expressions of the coefficients A,B,C,D in Eq.~\eqref{eq_A-D}. Let us recall that we have assumed that by the time the second photon is sent, the first photon has already been fully scattered by the QD, which means $\gamma, \Gamma \gg \Omega_g$: this also implies that in all the analytical expressions derived from the SPI wavefunction we put $e^{-\Gamma T_g/2} \rightarrow 0, e^{-\gamma T_g/2} \rightarrow 0$. In this condition, all quantities in the fidelity become functions of $\Gamma/\gamma, \Omega_{g,e}/\gamma$ and hence for ease of notation we set $\gamma = 1$ in the below expressions. Furthermore, the expressions of the overlaps $ \Lambda^{(\mu, \nu)}$ greatly simplify. For instance,
$\Lambda^{(\uparrow, \uparrow)}$ becomes a very simple function of the band-width ratio $\Gamma/\gamma$:
\begin{equation}\label{eq:Lambdaupup}
   \Lambda^{(\uparrow, \uparrow)} \rightarrow \frac{-1 + \Gamma^2}{\sqrt{2}(1 + \Gamma)^2}.
\end{equation}

 Using $n_x = 1$ also gives 
\begin{equation}
C \rightarrow\frac{
\begin{aligned}
&\{4 \Gamma^3 + 6 \Gamma^4 + \Gamma (\Gamma - \Omega_g) - 3 \Gamma^2 \Omega_g - 3 \Gamma^3 \Omega_g + \Omega_g^3 + \Omega_g^2 (\Omega_e^2 + \Omega_g^2) + \Gamma^2 (2 \Omega_e^2 + 3 \Omega_g^2) + \Gamma (-\Omega_e^2 \Omega_g + \Omega_g^3) \\
&+ (\Gamma^2 + \Omega_g^2) \left[ \Gamma^4 + (\Omega_e^2 - \Omega_g^2)^2 + 2 \Gamma^2 (\Omega_e^2 + \Omega_g^2) \right] + (\Gamma^2 + \Omega_g^2) \left[ 4 \Gamma^3 - \Gamma^2 \Omega_g - \Omega_e^2 \Omega_g + \Omega_g^3 + 2 \Gamma (2 \Omega_e^2 + \Omega_g^2) \right]\}
\end{aligned}
}{
(1 + 2 \Gamma + \Gamma^2 + (\Omega_e - \Omega_g)^2)(\Gamma^2 + \Omega_g^2)(1 + 2 \Gamma + \Gamma^2 + (\Omega_e + \Omega_g)^2)
}
\end{equation}

As mentioned in the main, the condition $A=B=C=D=e^{i\delta}$ giving fidelity equal to one, can be reached at $n_x = 1$, this gives $A = 1$ and hence $\delta=0$. The above expression for C, as expected goes to 1 in the limit $\Omega_{e}/\gamma\rightarrow0$ and $\Omega_{g}/\gamma\rightarrow0$:
\begin{equation}
    C \rightarrow \frac{\Gamma^6 + 4 \Gamma^5 + 6 \Gamma^4 + 4 \Gamma^3 + \Gamma^2}{\Gamma^2(1 + 2\Gamma + \Gamma^2)^2} = 1
\end{equation}
Thus we have $C \rightarrow 1$ as long as $\Omega_e\approx\Omega_g \ll \gamma, \Gamma$. Note that the relation between $\Gamma, \gamma$ is still undetermined.

Using the result of $A, C$ in $B, D$ gives 
\begin{equation}
    B \rightarrow \frac{1}{2} + \frac{1}{\sqrt{2}}(-\Lambda^{(\uparrow, \uparrow)} + \Lambda^{(\downarrow, \downarrow)} + i \Lambda^{(\uparrow, \downarrow)})
\end{equation}
\begin{equation}
\begin{aligned}
    D \rightarrow \frac{-1}{\sqrt{2}}(i \Lambda^{(\uparrow, \downarrow)} + \Lambda^{(\downarrow, \downarrow)} + \Lambda^{\uparrow, \uparrow}) + (\Lambda^{(\uparrow, \uparrow)})^2  = \frac{-1}{\sqrt{2}}(i \Lambda^{(\uparrow, \downarrow)} + \Lambda^{(\downarrow, \downarrow)} - \Lambda^{\uparrow, \uparrow}) - \sqrt{2} \Lambda^{(\uparrow, \uparrow)}+ (\Lambda^{(\uparrow, \uparrow)})^2
\end{aligned}
\end{equation}
Now assuming $B=1$ gives 
\begin{equation}
    D = -\frac{1}{2} - \sqrt{2} \Lambda^{(\uparrow, \uparrow)} + (\Lambda^{(\uparrow, \uparrow)})^2
\end{equation}
Demanding that $D=1$ yields
\begin{equation}
    \Lambda^{(\uparrow, \uparrow)} = -\frac{1}{\sqrt{2}} \text{ or } \frac{3}{\sqrt{2}}
\end{equation}
Using the explicit expression of   $\Lambda^{(\uparrow, \uparrow)}$ given in Eq.~\eqref{eq:Lambdaupup} we see that the only possible solution, as $\Gamma$ is always positive, is $-1/\sqrt{2}$ that is attained when $\Gamma/\gamma \ll 1$. This thus fixes the last part of the chain of inequalities.

\end{widetext}
\end{document}